\documentclass[a4paper]{cas-dc}                                                                                                                                                                                                                                                                       

\usepackage[numbers]{natbib}

\usepackage{xcolor}
\usepackage{hyperref}
\PassOptionsToPackage{bookmarks=false}{hyperref}
\hypersetup{
    colorlinks=true,
    linkcolor=blue,
    filecolor=blue,      
    urlcolor=blue,
    citecolor=blue
}

\usepackage[ruled,vlined,linesnumbered]{algorithm2e}
\usepackage{caption}
\usepackage{subcaption}

\usepackage{algpseudocode}
\usepackage{listings}
\usepackage{enumerate}

\definecolor{darkgreen}{rgb}{0,0.5,0}

\definecolor{grigiomoltochiaro}{gray}{0.97}
\definecolor{verde}{rgb}{0,1,0}
\newcommand{\define}{\triangleq}

\begin{document}
\let\WriteBookmarks\relax
\def\floatpagepagefraction{1}
\def\textpagefraction{.001}

\shorttitle{Decentralised Identity Federations using Blockchain}

\shortauthors{Mirza et~al.}

\title [mode = title]{Decentralised Identity Federations using Blockchain}                      



%
\author[1]{Mirza Kamrul Bashar Shuhan}
\ead{shuhan.mirza@gmail.com}
\ead[url]{shuhanmirza.com}

\address[1]{bKash Limited, Dhaka, Bangladesh}    

\author[2]{Syed Md. Hasnayeen}
\ead{rummanhasnayeen94@gmail.com}

\address[2]{Dynamic Solution Innovators Limited,
    Dhaka,Bangladesh}

\author[3]{Tanmoy Krishna Das}
\ead{tanmoykrishnadas@gmail.com}
\address[3]{Shahjalal University of Science and Technology, Sylhet, Bangladesh}

\author[4]
{Md. Nazmus  Sakib}
\ead{nsakib1017@gmail.com}


\address[4]{PriyoSys Limited, Dhaka, Bangladesh}

\author[5,6]
{Md Sadek Ferdous}
\ead{sadek.ferdous@bracu.ac.bd}
\ead[URL]{https://www.msferdous.info/}

\address[5]{BRAC University, Dhaka,Bangladesh}

\address[6]{Imperial College London,
London,United Kingdom}


\begin{abstract}
Federated Identity Management has proven its worth by offering economic benefits and convenience to Service Providers and users alike. In such federations, the Identity Provider (IdP) is the solitary entity responsible for managing user credentials and generating assertions for the users, who are requesting access to a service provider's resource. This makes the IdP centralised and exhibits a single point of failure for the federation, making the federation prone to catastrophic damages. The paper presents our effort in designing and implementing a decentralised system in establishing an identity federation. In its attempt to decentralise the IdP in the federation, the proposed system relies on blockchain technology, thereby mitigating the single point of failure shortcoming of existing identity federations.  The system is designed using a set of requirements  In this article, we explore different aspects of designing and developing the system, present its protocol flow, analyse its performance, and evaluate its security using ProVerif, a state-of-the-art formal protocol verification tool.

\end{abstract}






\maketitle

\section{Introduction}

We are living in a world that is rapidly undergoing a fundamental change: its different aspects are being digitally transformed. This is not just the internet of things (IoT) or mobile computing, digital transformations are happening to all societal systems- traffic, health, government, logistics, education, marketing, etc. Consequently, nowadays, more and more crucial service providers (SP) have opted to go for digital operations and require their users to register to their systems. This is required to ensure a personalised user experience for the user. To ensure security, users need to choose and later utilise a crucial attribute known as the \textit{credential}, with the password being the most widely used credential in the world \cite{Papathanasaki_2022}. With the growing number of service providers, managing each user's credentials is becoming a challenging task \cite{ferdous2013portable}.

A solution to this challenge has resulted in the creation of Identity Management Systems (IMS). Among many IMS, Federated Identity Management (FIM) is a popular IMS, particularly within Educational and Government settings. FIM facilitates the creation of Identity Federations, a trusted virtual boundary among different entities for sharing their resources \cite{5561611}. Within an identity federation, there is a central entity called Identity Provider (IdP). There could be multiple service providers (SPs) where each SP relies on the identity service provided by the IdP. Security Assertion Markup Language (SAML) is a major standard for creating and maintaining an identity federation. An identity federation provides a number of advantages, such as trusted service provisioning, Single-Sign-On (SSO) experience, reduced password management, improved role-based access, and other identity management activities \cite{IFpaper}, thus proving beneficial for both users and SPs. However, FIM suffers from a crucial issue: the IdP being the central component introduces a single point of failure. If the IdP ceases to function, the SPs of the identity federation cannot function properly. 

A blockchain is an immutable transaction ledger, maintained within a distributed network of peer nodes \cite{chowdhury2019comparative}. These nodes each maintain a copy of the ledger by applying transactions that have been validated by a consensus protocol, grouped into blocks that include a hash that binds each block to the preceding block. This increases the security and integrity of data. In addition, blockchain, due to its decentralised nature, had additional advantages such as distributed data sharing and data availability. In this article we explore a novel idea of integrating blockchain within the architecture of an identity federation as the functionalities of the IdP can be decentralised, thereby creating the notion of a decentralised identity federation. 

\noindent \textbf{Contributions.} The main contributions of the article are:

\begin{itemize}
    \item The elaboration of the idea of a blockchain-based decentralised identity federation.
    \item An architecture of the proposed system based on a rigorous threat model and requirement analysis.
    \item A Proof of Concept (PoC) implementation of the presented architecture using a state-of-the-art private blockchain network.
    \item Detailed performance analysis of the implemented PoC, showcasing its applicability.
    \item Rigorous security verification of the underlying protocol of the implemented PoC using ProfVerif, a state-of-the-art protocol verifier \cite{BlanchetFOSAD14, BlanchetSmythJCS18}.
\end{itemize}

\noindent \textbf{Structure.} We present a brief background on federated identity management, blockchain, and their different aspects in Section \ref{sec:background}. We present our proposal of a decentralised identity federation along with a threat modeling and requirement analysis in Section \ref{sec:proposal}. We discuss different components of the architecture of the system and its implementation details in Section \ref{sec:architecture}. The protocol flow of the PoC is illustrated in Section \ref{sec:proto}. In Section \ref{sec:analysis}, the performance of the developed PoC is evaluated against a number of blockchain network configurations and user loads. In Section \ref{sec:discussion}, we discuss how the proposed system has satisfied different requirements, formally verify the protocol  and discuss the advantages of the proposed system. Finally, we conclude in Section \ref{sec:conclusion}.

\section{Background}
\label{sec:background}
In this section, we briefly discuss different aspects of Federated Identity Management (Section \ref{subsec:fim}) and blockchain (Section \ref{subsec:blockchain}).

\subsection{Federated Identity Management}
\label{subsec:fim}

As per \cite{ferdous2014mathematical}, an entity is a physical or logical object which can be uniquely identified within a certain context with its own \textit{identity}. Federated Identity Management (FIM) can be considered a business and identity management model in which two or more trusted parties agree to an association through a technical contract to facilitate Identity Management. It also consists of functions and protocols to provide assurance regarding the identity of an entity (a user) for the purpose of authentication, authorisation, and service provisioning \cite{ferdous2015managing}. 

FIM enables a user to access restricted resources seamlessly and securely from different organisations in different Identity Domains. An identity domain is the virtual boundary, context, or environment in which a digital identifier is valid \cite{josang2007usability}. An identifier  within an identity domain is an attribute whose value can be used to uniquely identify a user within that identity domain. Examples of identifiers are username and email as their values can uniquely identify a user.

FIM offers a good number of advantages to different stakeholders such as the separation of duties among different organisations, scalability, improved security and privacy, Single Sign On (SSO) for users, and so on \cite{IFpaper}. For example, users can take advantage of SSO and thus authenticate themselves in one identity domain and receive personalised services across multiple domains without any further authentication. There are three major actors within an FIM System:
\begin{itemize}
    \item \textbf{Identity Provider (IdP):} An entity that is responsible for managing the digital identities of users and providing identity-related services to different Service Providers. IdPs are also known as Asserting Parties (AP). 
    \item \textbf{Service Provider (SP):} An entity that is responsible for providing online services to the users based  on  the identity  information  (identifiers  and/or  attributes)  received  from  the  IdP. SPs  are  also  known  as  Relying Parties (RP).
    \item \textbf{Users:} An entity that receives services from an SP.
\end{itemize}

\vspace{2mm}
\noindent \textbf{SAML-based FIM:} One crucial component of an identity federation is how trust is established among different organisations. SAML (Security Assertion Markup Language) is the most widely used technology for establishing trust among organisations and deploying identity federations among themselves \cite{samlv2, ferdous2013dynamic}. SAML is an XML-based standard for exchanging authentication and authorisation information between trusted yet autonomous organisational domains. It is based on the request/response protocol in which one party (generally SPs) requests particular identity information about a user and the other party ( IdPs) then responds with the information.

\begin{figure}
\centering
\includegraphics[width=\linewidth]{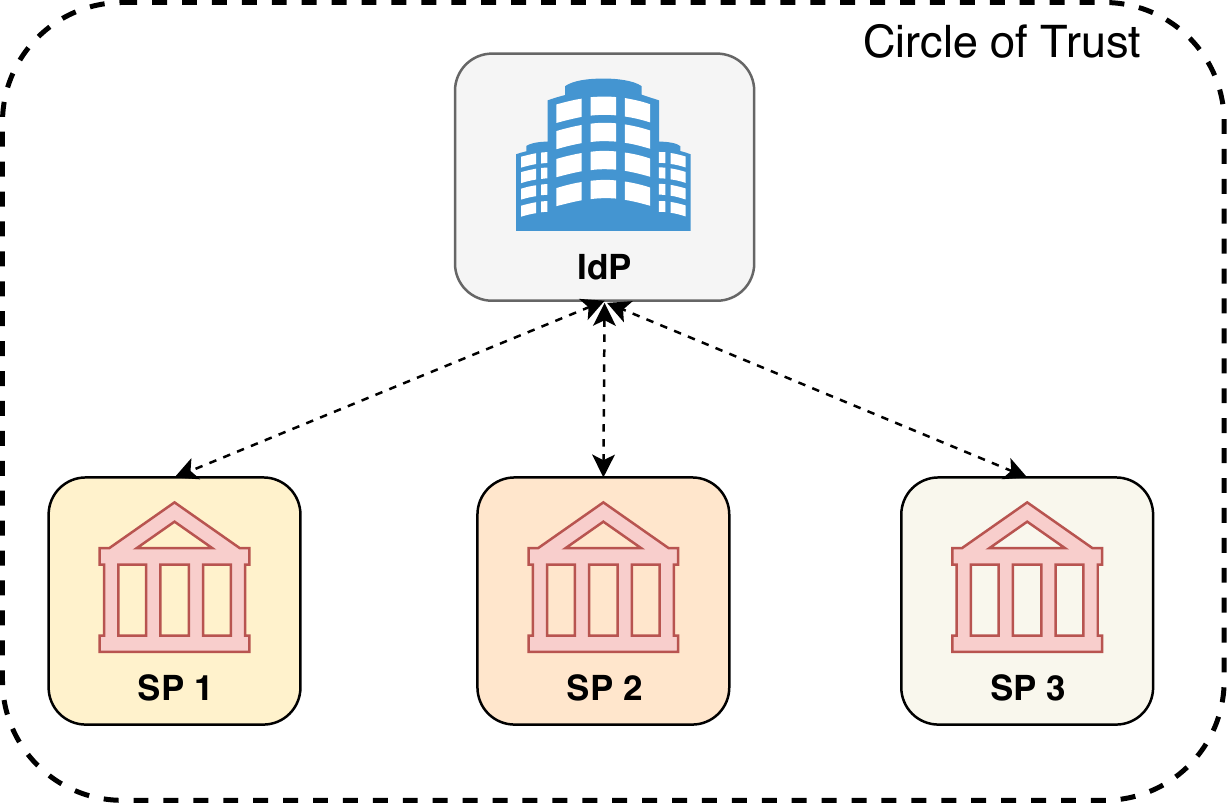}
\caption{SAML Identity Federation}
\label{Fig:samlFED}
\end{figure}

Metadata is a central component in a SAML-based identity federation and plays a crucial role in establishing trust among organisations. It is an XML file in a specified format containing several pieces of information such as entity descriptor (identifier for each party), service endpoints (the locations of the appropriate endpoints for IdPs and SPs), certificate(s) to be used for encryption, the expiration time of metadata, contact information, and other information. To establish a federation, IdPs and SPs exchange their metadata with each other and store them at the appropriate repositories at their ends which helps each party to build up the so-called Trust Anchor List (TAL). IdPs only trust those SPs whose metadata can be found in their TALs and vice versa, thus creating the notion of a trusted relationship, the so-called Circle of Trust (CoT), within FIM. Therefore, the TAL of an IdP consists of the metadata of SPs federated with the IdP and vice versa. Figure \ref{Fig:samlFED} represents a SAML identity federation where the dottet lines represent the CoT for the entities within that federation.

A SAML protocol flow between a user (denoted as $u$), an IdP (denoted as $\mathit{idp}$) and an SP (denoted as $\mathit{sp}$) is presented below using the notations presented in Table \ref{table:cryptoNotationSaml} and Table \ref{table:notationsSAML}  \cite{ferdous2020social}.

\begin{table}[h]
\caption{Identity, SAML \& Cryptographic Notations}
\label{table:cryptoNotationSaml}
\centering
\begin{tabular}{r|l}
\hline
\textbf{Notations}            & \textbf{Description} \\ \hline \hline
${id}_{sp|idp}$      & Entity Id of an of $sp$ or $idp$ \\ \hline
${\mathit{id}}_{req}$      & Identifier of an of a SAML request \\ \hline
$K_{sp|idp}$      & Public key of $sp$ or $idp$ \\ \hline
$K^{-1}_{sp|idp}$      & Private key of $sp$  or idp \\ \hline
${\{\}}_K$     & Encryption operation using $K$  \\ \hline
${\{\}}_{K^{-1}}$     & Signature using a private key $K^{-1}$  \\ \hline
$\mathit{AuthnReq}$ & SAML Authentication Request  \\ \hline
$\mathit{SAMLAssrtn}$ & SAML Assertion \\ \hline
$\mathit{EncSAMLAssrtn}$ & Encrypted SAML Assertion\\ \hline
$\mathit{SAMLResp}$ & SAML Response \\ \hline
\end{tabular}
\end{table}

\begin{table}[h]
\centering
\caption{SAML notations}
\label{table:notationsSAML}
\begin{tabular}{l}
\hline
$\mathit{AuthnReq} \define \langle\ {id}_{req}, {id}_{sp} \, \,\rangle $\\
$\mathit{SAMLAssrtn} \define \langle\ \mathit{PROFILE}_{idp}^u  \, \,\rangle $\\
$\mathit{EncSAMLAssrtn} \define \langle\ {\{\mathit{SAMLAssrtn}\}}_{K_{sp}} \, \,\rangle $ \\
$\mathit{Assrtn} \define \langle\ {\{\mathit{SAMLAssrtn}\}}_{K^{-1}_{idp}} | {\{\mathit{EncSAMLAssrtn}\}}_{K^{-1}_{idp}}\, \,\rangle$ \\
$\mathit{SAMLResp} \define \langle\ {id}_{req}, {id}_{sp}, {id}_{idp}, \mathit{Assrtn} \, \,\rangle $ \\\hline
\end{tabular}
\normalsize
\end{table}

While trying to access a service provided by \textit{sp}, \textit{u} is forwarded to a special service called the \textit{Discovery Service} or \textit{Where Are You From (WAYF)} Service. The WAYF shows a list of pre-configured trusted IdPs ($\mathit{IDP_{wayf}}$) to $u$. After choosing her preferred IdP, $u$ is forwarded to \textit{idp} with a SAML authentication request consisting of an identifier of the request and an identifier (called \textit{entity ID} in SAML) of $sp$. A SAML request is denoted using $\mathit{AuthnReq}$ and is modelled as presented in Table \ref{table:notationsSAML}, where ${id}_{req}$ representing the identifier in each SAML request and ${id}_{sp}$ denoting the entity ID of $sp$. At $\mathit{idp}$, $u$ is authenticated at first, and then $\mathit{idp}$ prepares a SAML response with an embedded SAML assertion. The assertion contains the user profile (explained below) as released by $idp$. Mathematically, the assertion is denoted with $\mathit{SAMLAssrtn}$ and is modelled as per Table \ref{table:notationsSAML}. 

Then, $\mathit{idp}$ digitally signs the (encrypted or unencrypted) SAML assertion and then embeds it inside a SAML response. The response also contains the request identifier (${id}_{req}$), the entity ID (${id}_{idp}$) of $\mathit{idp}$ and the entity ID (${id}_{sp}$) of $sp$. A SAML response is denoted with $\mathit{SAMLResp}$ and modelled as per Table \ref{table:notationsSAML}.

Finally, the response is sent back to $sp$ by the $\mathit{idp}$. Upon receiving the response, the (encrypted/unencrypted) SAML assertion is extracted. If the response consists of an encrypted assertion, $sp$ decrypts the assertion at first with the private key of $sp$ ($K^{-1}_{sp}$) and then validates the signature with the public key of $\mathit{idp}$ ($K_{idp}$). In case of an unencrypted assertion, $sp$ just validates its signature using the public key of $\mathit{idp}$. $sp$ retrieves the embedded user attributes  from the assertion, only if the signature is valid, otherwise, the assertion is discarded. To ensure privacy, $\mathit{idp}$ does not release all the attributes to $sp$, instead, a subset of attributes are released. Such attributes are regarded as the profile of a user at $idp$, denoted with the notation ${\mathit{PROFILE}}^u_{\mathit{idp}}$. Consequently,  a SAML assertion is modelled to consist of the profile in Table \ref{table:notationsSAML}.

\vspace{2mm}
\noindent \textbf{Issues in FIM:} One crucial bottleneck within a federation is that it is centralised. There are two major implications of this issue. The first implication is that if the IdP within a SAML Identity federation malfunctions, the users of that federation will not be able to access federated services from that domain, thereby exhibiting a single point of failure. The second reason is that, if the IdP within a federation relies on a centralised database to store user credentials and attributes and such storage server does not work properly, the IdP will not be able to provide required identity services to the SPs within the federation, thereby, causing disruptions towards federated services. In this work, we present a novel blockchain-based approach to tackle both these centralisation issues.

\subsection{Blockchain}
\label{subsec:blockchain}
Because of the zero-reliance on any central entity such as a central bank, Bitcoin is often considered as the first successful decentralised digital currency \cite{nakamoto2019bitcoin}. The technological innovation of Bitcoin is underpinned by a smartly-engineered solution known as \textit{blockchain}. A blockchain is essentially a distributed ledger consisting of transactions which are grouped together using the concept of blocks and these blocks are chained consecutively following a strict set of rules \cite{chowdhury2019comparative}. Each transaction and block are consequently verified by many distributed Peer-to-Peer (P2P) nodes, thereby ensuring that the system can function even in midst of attacking/corrupting P2P nodes which may not follow the protocol rules properly. Each transaction in the blockchain represents an instruction to transfer value or data from one entity to another. Blockchain offers a number of advantages such as data immutability, data provenance, distributed data sharing and so on. Evolving from Bitcoin, another generation of blockchain system has emerged which supports the deployment of smart-contracts on top of the respective blockchain platform. A smart-contract (SC) is a computer program which can be executed by a computing platform that is integrated with a blockchain system. Being rooted on blockchain, the code of an SC and the data it stores become an integrated part of an immutably ledger, thereby facilitating the notion of immutable logic, which is a sort after property in many application domains \cite{ferdous2019search}. A blockchain can be \textit{public}, allowing everyone to participate, or \textit{private}, where only authorised parties can participate. Bitcoin \cite{bitcoin2018} and Ethereum \cite{ethereum2018} are examples of public blockchains, whereas Hyperledger platforms \cite{hyperledger2018} and Quoram \cite{quorum2018} are examples of private blockchain systems.

\section{Proposal, Threat Modelling \& Requirements Analysis}
\label{sec:proposal}
In this section, we present our proposal (Section \ref{sec:PTR:subset:tm}) and a threat model (Section \ref{subsec:threat}) and then analyse many functional and security requirements (Section \ref{subsec:req}) for the proposed blockchain-based framework for a decentralised identity federation.

\subsection{Proposal}
\label{sec:PTR:subset:tm}
To mitigate the identified issue of a single point of failure in an identity federation, we propose to devise a mechanism that will essentially decentralise the functionalities of an IdP.  Towards that aim, we propose to disrupt the current setting of an identity federation by introducing a blockchain-based `inner' federation of many IdPs within a single identity federation. These inner groups of federated IdPs within another single federation combinedly act like a single IdP to any SP within the federation. This novel proposal will require to make changes regarding how an identity federation is established and maintained, and its services are provisioned.  Before explaining how we have achieved these goals, we present the threats and the requirements for such a system in the following sections. 

\subsection{Threat Modelling}
\label{subsec:threat}
Threat modeling is a crucial step for designing and developing a secure framework for mitigating threats involving IT assets, identity federations in the scope of this paper. To model threats, we have chosen a well-established threat model called STRIDE \cite{shostack2014threat}, which encapsulates six security threats. Next, we discuss five of these six threats modelled within the scope of the current work. The last threat  (the `E' threat in STRIDE which implies \textit{Elevation of Privilege}) is excluded as it is related to an authorisation which is generally carried out by an SP once it receives an assertion from an IdP and thus, it is beyond the scope of the current work. 

\begin{itemize}
    \item \textbf{T1-Spoofing:} An entity (e.g. SP) can pretend to provide a federated service even though it is not part of any federation. 
  
    \item \textbf{T2-Tampering:} An entity (e.g. SP or IdP) might alter the TAL of another entity to be in the CoT without exchanging the required metadata. 
    
    \item \textbf{T3-Repudiation:} An IdP can repudiate that it has not released any assertion to an SP. 
    \item \textbf{T4-Information Disclosure:} We consider two different types of T4 threat: 
	\begin{itemize}
			\item \textbf{T4-1:} User attributes are disclosed to an unauthorised attacker.
			\item \textbf{T4-2:} User attributes from an IdP are disclosed to an SP without the user's knowledge or consent.
		\end{itemize}
    \item \textbf{T5-Denial of Service (DoS):} The federated services become unavailable because of an entity (an IdP) being unavailable due to a DoS attack. 
\end{itemize}
Apart from STRIDE threats, we also consider the following additional threat.
\begin{itemize}
    \item \textbf{T7- Replay Attack:} An attacker can capture and reuse any previous SAML packet (request, response, or assertion) for any malicious intent.
\end{itemize}

\subsection{Requirement Analysis}
\label{subsec:req}

Accurate and well-defined requirement analysis is an essential part of any successful application or system development process. Before implementing our proposed approach, we formulated many functional and security (non-functional) requirements. The functional requirements represent the absolutely necessary features of the proposed approach whereas the security requirements are needed to ensure the security of the approach so as to mitigate the identified security threats. Next, we present the requirements. 

\subsection{Functional Requirements}
At first, the functional requirements are presented
\begin{itemize}
    \item \textbf{F1}: There should be a mechanism to establish a trusted relationship among different IdPs within another identity federation.
    \item \textbf{F2}: Such IdPs should have the provision to share the attributes of their users with other trusted IdPs.
    \item \textbf{F3}: In case an IdP ceases to function, a mechanism should be established for a user to select one of the other federated IdPs. 
    \item \textbf{F4}: The functionalities should be integrated in such a way that it has a minimal impact on any existing components of SAML based system.

\end{itemize}

\subsection{Non-Functional (Security) Requirements}
Next, we present the security requirements of the system.
\begin{itemize}
    \item \textbf{S1}: The usual trust requirements of any SAML federation are ensured. This trust requirement ensures that only federated entities can request and avail identity and avail federated services within a federation, hereby mitigating T1 and only trusted entities can exchange their metadata and store them in their respective TAL, thereby mitigating T2.
    
    \item \textbf{S2}: Every released assertion must be digitally signed so that the IdP cannot repudiate. This mitigates T3.
    \item \textbf{S3}: An IdP should always respond with an encrypted assertion (\textit{EncSAMLAssrtn}) for the requesting SP. In general, if all data transmission is carried out in an encrypted channel (HTTPS), then the encrypted assertion requirement can be relaxed. Any of these will mitigate T4-1. 
    \item \textbf{S4}: Every user attribute should be released only after the respective user has provided their explicit consent to release such attribute. This will deter threat T4-2.
    \item \textbf{S5}: An SP within a federation should be able to avail the identity service of an IdP even if it becomes unavailable due to an attack such as DoS. This will mitigate threat T5.
\end{itemize}

\section{Architecture and Implementation}
\label{sec:architecture}
In this section, we present the architecture of the proposed system and discuss how the architecture has been implemented.

The architecture of the proposed system is presented in Figure  \ref{Fig:coreArchi} and Figure \ref{Fig:combinedCoT}.  Next, we explain the functionalities of different components of this architecture.

\subsection{\textbf{Combined IdP}}
\begin{figure}
\includegraphics[width=0.8\linewidth]{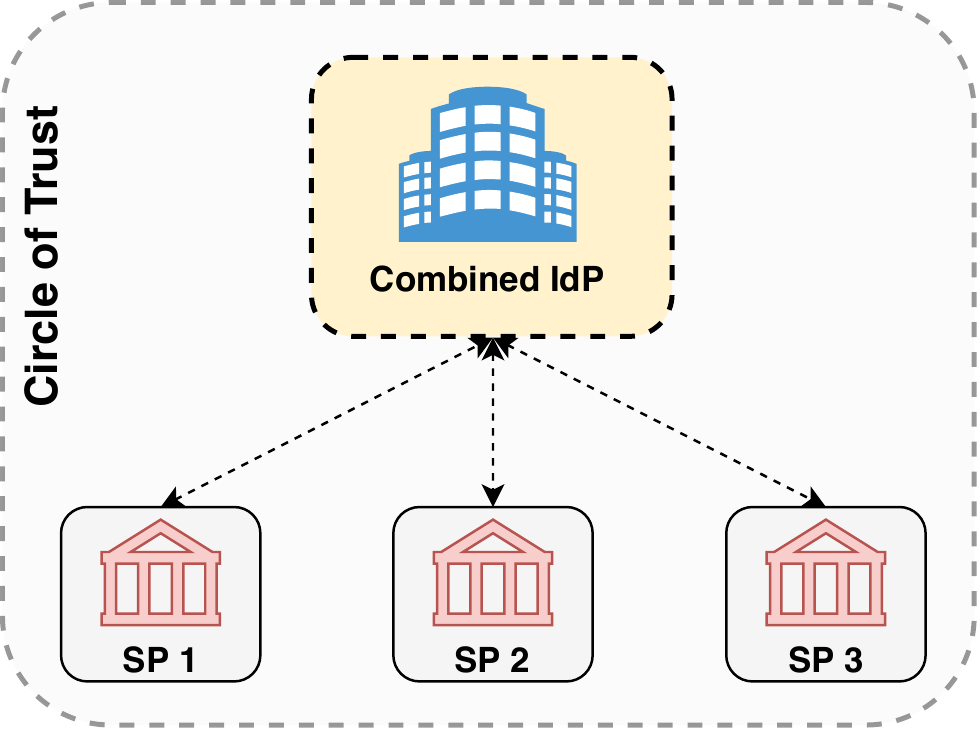}
\centering
\caption{Decentralised Identity Federation Architecture}
\label{Fig:coreArchi}
\end{figure}

\begin{figure}
\includegraphics[width=0.8\linewidth]{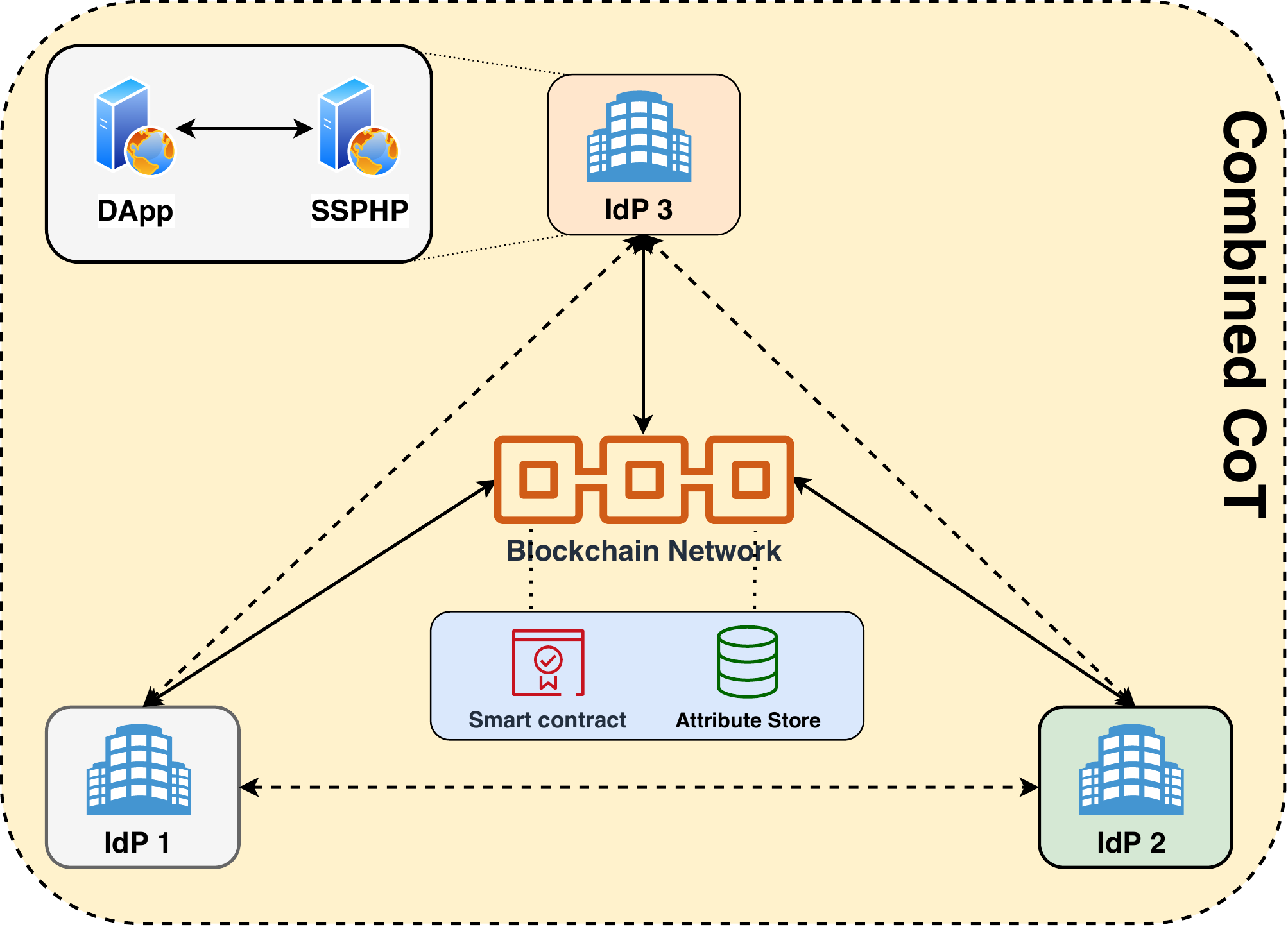}
\centering
\caption{Individual Components inside Combined CoT}
\label{Fig:combinedCoT}
\end{figure}

Our proposal evolves around the idea that a number of IdPs are integrated in such a way that they act like a single IdP to all other SPs within the federation.  This is illustrated in Figure \ref{Fig:coreArchi} where the integrated IdPs are denoted as the \textit{Combined IdP}. These IdPs are inter-connected with each other using a blockchain platform (Figure \ref{Fig:combinedCoT}). The dotted circle around the Combined IdP indicates that these integrated IdPs essentially form an implicit and combined Circle of Trust (Figure \ref{Fig:combinedCoT}) even though they are not federated with each other in a traditional way (e.g. metadata exchange).  Even though we have shown three IdPs in Figure \ref{Fig:coreArchi} and Figure \ref{Fig:combinedCoT}, we can add a few more IdPs to offer better availability of IdP services.

Each IdP has two sub-components (Figure \ref{Fig:coreArchi}): SSPHP (SimpleSAMLPHP) and DApp (Decentralised Application). There are several implementations of the SAML standard, such as Shibboleth \cite{shibboleth}, SimpleSAMLphp \cite{ssphp} and ZXID \cite{zxid}. Among them, we have used SimpleSAMLPHP (\textit{SSPHP} in short), a SAML implementation developed in PHP, for our implementation as it is light-weight, easy to deploy, has additional extensibility using the concept of modules and is fully open-sourced. We have modified the code base of the SSPHP to meet our requirements (explained subsequently). Each of the combined IdPs is attached to this modified SSPHP instance which takes care of most of the SAML functionalities. 

A DApp is a web server that acts as the middleware between our SAML IdPs and a blockchain. Since a SAML-based IdP does not have any provision for interacting with the blockchain, we have utilised a DApp, each for one IdP, to interact with the blockchain platform. Each of these DApps exposes APIs to the respective SSPHP instance of the IdP and is also connected to a peer of the blockchain platform. The SAML interface of each IdP uses these APIs to submit some requests (e.g. for some specific purposes explained later) to the DApp. These requests are translated into blockchain transactions by the DApp for submission to the blockchain platform using the blockchain API via a peer. 

The DApp in the proposed framework has been developed using Node.js with Express \cite{nodejs, express}. Node.js is a server-side JavaScript platform that is widely used for creating DApps in the blockchain domain. Express is a web application framework for Node.js, which is used for developing web applications.

\subsection{\textbf{Blockchain Platform}}

The blockchain platform plays a crucial role to achieve the desired functionalities of the proposed system. The blockchain platform serves the following three purposes within the proposed framework:

\begin{itemize}
     \item Being rooted on top of a decentralised blockchain platform facilitates the provision to establish the notion of trust among the integrated IdPs.
     \item The distributed data-sharing nature of blockchain provides the underlying mechanism to share the user credentials and other attributes of among the IdPs in a timely and synchronised way. This implies that the blockchain will be used as an attribute store as well. 
    \item The smart contract (SC) in the blockchain is used to encode immutable logic for managing the trust-relationship between the IdPs and storing attributes and credentials. 
\end{itemize}

An SC supported public blockchain (e.g., Ethereum) is relatively slow, open to all and  expensive to process and store data   \cite{chowdhury2019comparative}. On the other hand, private blockchain systems are private and fast with no issue of energy consumption and provide a reasonable amount of security. Furthermore, a federation is essentially a closed network of trusted entities that highlights the requirement for a private blockchain platform. For these reasons, we have utilised a private blockchain platform.

Among many private blockchain platforms, Hyperledger Fabric is currently the most stable and popular private blockchain platform \cite{bhuiyan2020bonik}. It is also equipped with a unique concept of \textit{channel} which allows different fabric blockchains to be maintained within the same network, thus creating a privacy layer among different organisations. That is why Hyperledger Fabric is selected as our preferred blockchain platform during the deployment phase. 

The fabric utilises many network entities such as peers, endorsers, and orderers. A smart contract is called a \textit{chaincode} in Fabric terminology which can be invoked using transactions. Transactions are submitted via peers, endorsers are responsible for validating transactions and orderers create blocks.

The chaincode for the proposed framework has been written in Go. The Fabric network contains two organisations, each representing an IdP and an SP. Each organisation consists of two peers/endorsers. We have used Docker containers to deploy the platform with this network configurations. An additional entity called MSP (Membership Service Provider), including a CA (Certificate Authority), is also deployed with another container. All nodes are connected via a channel. We have utilised Kafka (a distributed event streaming platform \cite{Kafka}) consensus algorithm with two additional orderer nodes for block creation and dissemination. 

\vspace{3mm}
\noindent \textbf{Service Provider (SP):}
The Service Providers (SPs) in the proposed architecture function mostly similarly to any traditional SAML SP.  However, instead of being federated with a single IdP in the federation, each IdP is federated with all of the combined IdPs. This is used to bootstrap the trust relationship between the combined IdPs and other federated SPs. How it is carried out and the protocol flow involving an SP and the combined IdPs will be explained in Section \ref{sec:proto}. To accommodate this protocol flow, we have modified the SP-side code-based of SimpleSAMLPHP as well.

\section{Protocol Flow \& Use-case}
\label{sec:proto}
In this section, we present the protocol flow between different components of the proposed system. Before we illustrate the protocol flow, we introduce the mathematical notations in Table \ref{table:cryNot} and the data model in Table \ref{table:dModel}.

\subsection{Data Model}

\begin{table}[t]
\caption{Cryptographic Notations for Protocol Flow}
\label{table:cryNot}
\centering
\begin{tabular}{r|l}
\hline
\textbf{Notations}            & \textbf{Description} \\ \hline \hline

$\mathit{sp_i}$      & A service provider in the federation  \\ \hline
$\mathit{idp_i}$      & An identity provider in the federation  \\ \hline
${id}_{sp_i|idp_i}$      & Entity Id of an $sp_i$ or $idp_i$ \\ \hline
$CID$      & Common Entity ID for the combined $CoT$ \\ \hline
$A$      & Admin of the federation  \\ \hline
$K$      & A symmetric encryption key  \\ \hline
$K_{A}$      & Public key of $A$ \\ \hline
$K^{-1}_{A}$      & Private key of $A$ \\ \hline
$K_{D}$      & Public key of DApp \\ \hline
$K^{-1}_{D}$      & Private key of DApp \\ \hline
$N_i$    & A fresh nonce   \\ \hline
$H(.)$     & A hash function \\ \hline
${[ ]}_{\mathit{https}}$     & Communication over an HTTPS channel  \\ \hline
$\mathit{IDPList}$ & List of currently active IdPs  \\ \hline
$\mathit{AttList}$ & List of user attributes and their values \\ \hline
$a_i$ & A attribute name  \\ \hline
$v_{a_i}$ & Corresponding value for attribute $a_i$ \\ \hline
$\mathit{msg}$ & A textual message \\ \hline
$\mathit{B}$ & Fabric Blockchain Platform \\ \hline
$\mathit{CC}$ & Fabric Chaincode \\ \hline
\end{tabular}
\end{table}

\begin{table}[h]
\caption{Data Model}
\label{table:dModel}
\centering
\begin{tabular}{p{8cm}}
\hline
$\mathit{req} \define \langle type, data \rangle$\\ \hline

$\mathit{TYPE} \define \langle \mathit{idpReg, idpQuery, userReg, authn, login, cid} \rangle$\\ \hline
$\mathit{DATA} \define \langle \mathit{idpRegData, idpQData, userRegData, AuthnReq,}$ \newline $ \mathit{loginData} \rangle$\\ \hline
$\mathit{IDPList} \define \langle \mathit{{id}_{{idp}_1}, {id}_{{idp}_2}, ..., {id}_{{idp}_n}} \rangle$\\ \hline
$\mathit{idpRegData} \define \langle \mathit{{id}_{{idp}_i}, {\{{id}_{{idp}_i}\}}_{K^{-1}_{A}}} \rangle$ \\ \hline
$\mathit{idpQData} \define \langle \mathit{CID} \rangle$ \\ \hline
$\mathit{userRegData} \define \langle \mathit{userName, h, {\{AttList\}}_K} \rangle$ \\ \hline
$\mathit{AttList} \define \langle {\{(a_1, v^a_1), (a_2, v^a_2),...,(a_n, v^a_n)\}}_{K_{\mathit{idp}_i}} \rangle$ \\ \hline
$\mathit{loginData} \define \langle \mathit{userName, h} \rangle$ \\ \hline

$\mathit{resp} \define \langle msg, \mathit{SAMLResp}, \{msg||\mathit{SAMLResp}\} _{K^{-1}_{idp}} \rangle$\\ \hline
\end{tabular}
\end{table}

We start with the request (denoted with req
in Table \ref{table:dModel}), which is submitted to the blockchain platform. req consists of type and data. Here, TYPE denotes the set of different data types within a request and $\mathit{type} \in \mathit{TYPE}$, whereas, DATA represent the set of corresponding data and $\mathit{data} \in \mathit{DATA}$. Both TYPE and DATA are defined as presented in Table \ref{table:dModel}.

There are two types of registration requests in the system: IdP registration ($\mathit{idpReg}$) and User Registration  ($\mathit{userReg}$). $\mathit{idpReg}$ signifies the registration of IdPs within the combined IdP set and contains the entity ID of the IdP ($\mathit{{id}_{{idp}_i}}$) and a digital signature as defined with $\mathit{idpRegData}$ in Table \ref{table:dModel}.

On the other hand, the $\mathit{userReg}$ signifies that the corresponding request will be a user registration request to an IdP consisting of the data set denoted with $\mathit{userRegData}$ in Table \ref{table:dModel}. In $\mathit{userRegData}$, $h = H(\mathit{Password})$ denotes the hash of the provided password, $\mathit{userName}$ denotes the username (identifier) of the user and $\mathit{AttList}$ contains the additional attribute and their values as required during the registration. This implies that a registration request must contain a username, the hash of the password, and additional attributes of the user. $\mathit{loginData}$ also has a similar semantic in the sense that a login request must consist of the username and the hash of the provided password.

The $\mathit{idpQuery}$ type denotes a special type of request to retrieve a specific IdP from the combined set of IdPs. The corresponding data for this type is $\mathit{idpQData}$ which consists of the common entity ID (denoted with $\mathit{CID}$) for the combined set of IdPs. The motivation for utilising such a common entity ID for the combined IdPs is as follows. As mentioned earlier, each IdP within SAML will have a separate entity ID ($\mathit{{id}_{{idp}_1}}, \mathit{{id}_{{idp}_2}}, \mathit{{id}_{{idp}_3}}$ and so on). Utilising all these IdPs might confuse the user. To improve the user experience, the combined IdPs will be externally denoted with $\mathit{CID}$ even though internally they will be identified with their respective entity ID. The $\mathit{idpReg}$ will be used to bind different IdPs within a common entity ID and the $\mathit{idpQuery}$ is used to retrieve the binding entity IDs under a common entity ID. 

Additionally, $\mathit{authn}$ represents an authentication request whose data is essentially a SAML authentication request (denoted with $AuthnReq$ in Table \ref{table:notationsSAML}). 

A response with respect to a particular request type is denoted with $\mathit{resp}$ where a response contains a textual message ($msg$), a SAML response ($\mathit{SAMLResp}$, as defined in Table \ref{table:notationsSAML}) and a signature of the concatenated $msg$ and $\mathit{SAMLResp}$ by the IdP.

\subsection{Protocol Flow}
In this section, we present the protocol flows involving different entities of the architecture. The protocol flows are divided into three phases: i) IdP Registration \& setup phase, ii) User registration phase, and ii) Service provisioning phase. 

\subsubsection{Registration \& setup phase} This is the first step towards creating a decentralised federation where three IdPs are merged as a combined IdP (CoT). An admin of the organisation is assumed to take the responsibility for creating the federation. At first, the admin uses the SimpleSAMLPHP framework to set up three IdPs within the same organisation. These IdPs have three different entity IDs and are deployed in three different hosts. Then, these IdPs are connected to the same blockchain network so that they can share the same blockchain. To facilitate this, a DApp is deployed at each IdP. The SimpleSAMLPHP code for each of these IdPs has been modified so that it can interact with their corresponding DApp and via the DApp with the Fabric blockchain component. This ensures that the SimpleSAMLPHP can be utilised for this as well as for the protocol flows for the other two phases. 

\begin{algorithm}[h]
\SetAlgoLined
\caption{Chaincode} \label{algo:scc}
\textbf{Input:} $req \rightarrow$ the request from the user \\
\textbf{Output:} $resp \rightarrow$ the chaincode generated response\\
\SetKwBlock{Begin}{}{}
\Begin(\textbf{Start})
{
  $CID \leftarrow $ Generate CID \;      
  \SetKwProg{invoke}{function \textbf{invoke}}{($req$)}{}
  \invoke{}{
   $data \leftarrow req.data$\;
   $type \leftarrow req.type$\;
  \uIf{type = idpReg}{
    $resp \leftarrow $ regIDP($data$)\; 
  }
  \uElseIf{type = idpQuery}{
    $resp \leftarrow $ queryIDP($data$)\; 
  }

  \uElseIf{type = userReg}{
    $resp \leftarrow $ regUser($data$)\; 
  }
  
  \uElseIf{type = login}{
    $resp \leftarrow $ loginUser($data$)\; 
  }
  
  \uElseIf{type = cid}{
    $resp \leftarrow CID$\; 
  }
  
   send $resp$ back to DApp\;
  }
  \SetKwProg{queryIDP}{function \textbf{queryIDP}}{($data$)}{}
  \queryIDP{}{
  \textit{idps} $\leftarrow$ \textit{getState(CID)}\;
  \KwRet \textit{idps}\;
  }
  
  \SetKwProg{regIDP}{function \textbf{regIDP}}{($data$)}{}
  \regIDP{}{
  $idp \leftarrow data.{id}_{{idp}_i}$\;
  \textit{idps} $\leftarrow$ \textit{getState(CID)}\;
  
  \uIf{idp $\notin$ idps}{
    \textit{IDPList} $\leftarrow$ \textit{IDPList} $\cup$ \textit{idp} \;
    \textit{putState(CID,IDPList)}\;
    \KwRet \textit{TRUE}\;
  }\Else{
    \KwRet \textit{FALSE}\;
  }
  }
  
  \SetKwProg{regUser}{function \textbf{regUser}}{($data$)}{}
  \regUser{}{
  \textit{userName} $\leftarrow$ \textit{data.userName}\;
  \textit{hash} $\leftarrow$ \textit{data.h}\;
  \textit{AttList} $\leftarrow$ \textit{data.AttList}\;
  
  \textit{putState(userName,(hash,AttList))}\;
  
  \KwRet \textit{TRUE}\;
  }
  
  \SetKwProg{loginUser}{function \textbf{loginUser}}{($data$)}{}
  \loginUser{}{
  \textit{userName} $\leftarrow$ \textit{data.userName}\;
  \textit{hash} $\leftarrow$ \textit{data.h}\;
  \textit{ledgerData} $\leftarrow$ \textit{getState(userName)}\;  
  $h \leftarrow$ \textit{ledgerData[0]}\;
  \uIf{$h == hash$}{
    \KwRet \textit{ledgerData[1]}\;
  }\Else{
    \KwRet \textit{FALSE}\;
   }
  }
}
\end{algorithm}

To set up the Combined IdP, the smart-contract (chaincode) of the blockchain component is utilised. The algorithm for the chaincode is presented in Algorithm \ref{algo:scc}. The entry point for the chaincode is the \textit{invoke} function. Every request transmitted via the DApp is transformed into a transaction and is intercepted by the invoke function. Depending on the type of the request, the invoke functions calls different functions and the response from the calling function is sent back to the DApp (line 4 to 11 in Algorithm \ref{algo:scc}). For example, if the type of the request is \textit{idpReg}, which signifies registering an IdP to the combined CoT, then \textit{regIDP} function is called (line 7-8 in Algorithm \ref{algo:scc}). Similarly, to check if an IdP is part of the combined CoT, the request will contain an \textit{idpQuery} type. In this case, the invoke function will call thee \textit{idpQuery} function.

\begin{figure*}
\includegraphics[width=0.8\linewidth]{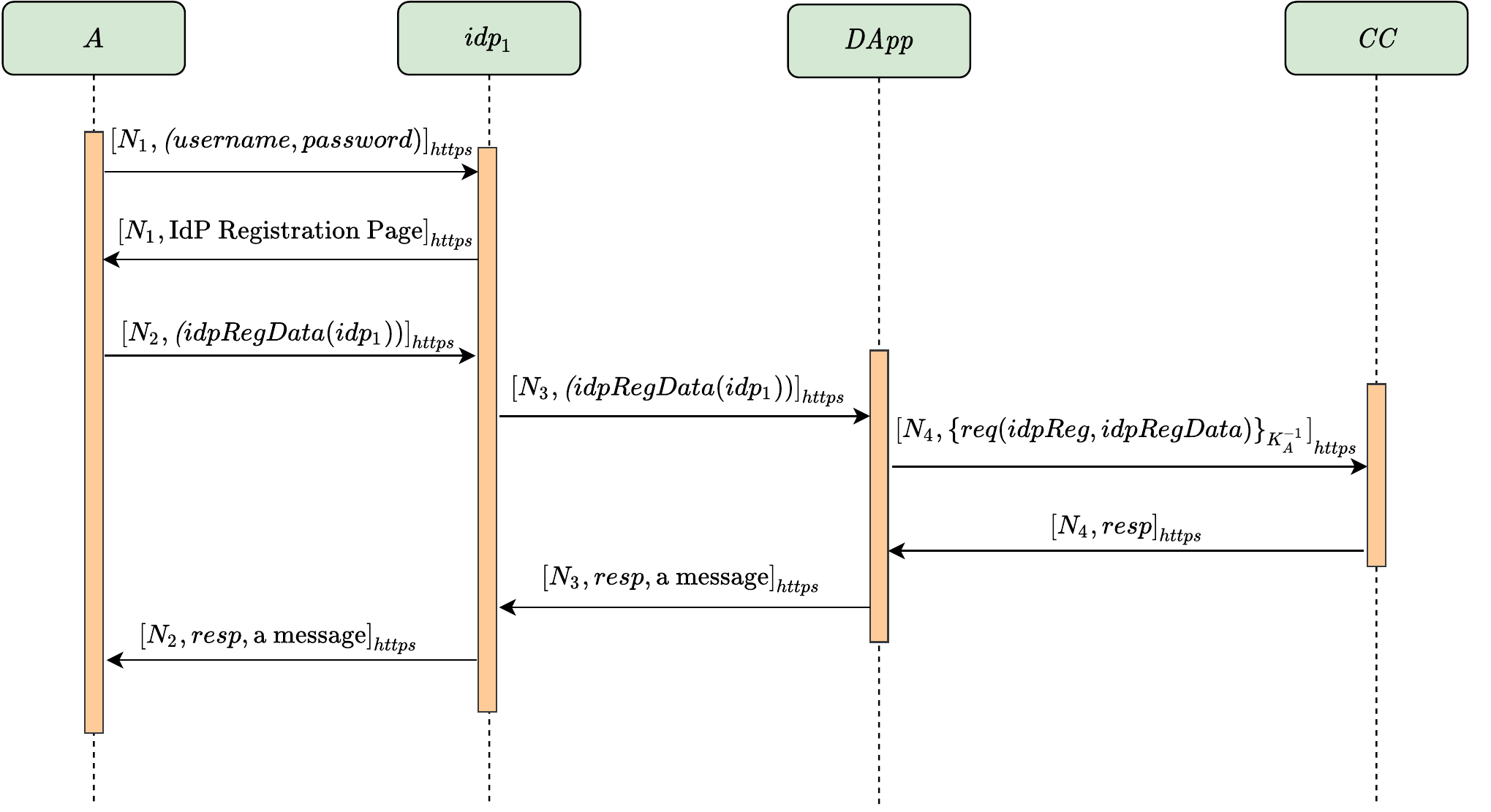}
\centering
\caption{Protocol Flow for IdP registration and setup}
\label{Fig:proto1}
\end{figure*}

Next, the admin engages in the following protocol flow to create the notion of the combined CoT out of these three IdPs. The flow is illustrated in Figure \ref{Fig:proto1} and its corresponding protocol is presented in Table \ref{table:regProtocol}.

\begin{table}[h]
\caption{Registration \& setup protocol}
\label{table:regProtocol}
\centering
\begin{tabular}{p{0.2cm}lp{4.5cm}}
\hline
$M1$ & $A \rightarrow \mathit{idp}_1:$  & $[N_1, \mathit{username, password}]_{\mathit{https}}$ \\
$M2$ & $\mathit{idp}_1 \rightarrow A:$  & $[N_1, \text{IdP Registration Page}]_{\mathit{https}}$ \\
$M3$ & $A \rightarrow \mathit{idp}_1:$  & $[N_2, \mathit{idpRegData} \text{ with } \mathit{id}_{\mathit{idp}_1}]_{\mathit{https}}$ \\
$M4$ & $\mathit{idp}_1 \rightarrow \mathit{DApp}:$  & $[N_3, \mathit{idpRegData} \text{ with } \mathit{id}_{\mathit{idp}_1}]_{\mathit{https}}$ \\
$M5$ & $\mathit{DApp} \rightarrow CC:$  & $[N_4, \{\mathit{req(idpReg, idpRegData)\}_{K^{-1}_{A}}}]_{\mathit{https}}$ \\
$M6$ & $CC \rightarrow \mathit{DApp}:$  & $[N_4, \mathit{resp}]_{\mathit{https}}$ \\
$M7$ & $\mathit{DApp} \rightarrow \mathit{idp}_1:$  & $[N_3, \mathit{resp} \text{ with a meaningful message}]_{\mathit{https}}$ \\
$M8$ & $\mathit{idp}_1 \rightarrow A:$  & $[N_2, \mathit{resp} \text{ a meaninfgul message}]_{\mathit{https}}$ \\
\hline
\end{tabular}
\end{table}

\begin{enumerate}[i]
    \item Each IdP has a restricted service for registering the IdPs. The service can be accessed only by the admin of the IdP. The admin logs in to that restricted service of one of the IdPs (let us assume that it is denoted with $\mathit{idp}_1$). This is represented in $M1$ and $M2$ steps in Table \ref{table:regProtocol}.
    \item The admin provides the entity ID of the logged in IdP and clicks the \textit{Register} button ($M3$ in Table \ref{table:regProtocol}).
    \item This information is submitted to the corresponding DApp of the IdP ($M4$ in Table \ref{table:regProtocol}).
    \item The DApp creates and signs a transaction consisting of an IdP registration request. This request contains the the type as \textit{idpReg} and data as \textit{idpRegData} as defined in Table \ref{table:dModel}. This request embedded within the transaction is then submitted to the blockchain to initiate the Fabric flow.
    \item Once the transaction is approved by the endorsers of the blockchain platform, it is forwarded to the invoke function of the chaincode ($M5$ in Table \ref{table:regProtocol}).
    \item The invoke function checks the type of the request and as it is an \textit{idpReg} request, it is forwarded to the \textit{regIDP} function (line 6 to 9 in Algorithm \ref{algo:scc}).
    \item The \textit{regIDP} function firstly retrieves the entity ID of the IdP from the request (line 23 in Algorithm \ref{algo:scc}).
    \item Then, the \textit{regIDP} function retrieves the list of registered IdPs against the entity ID of the Combined IdP from the blockchain (line 24 in Algorithm \ref{algo:scc}). The entity ID of the combined IdP is denoted with CID in the algorithm and is generated when the chaincode is initiated for the first time (line4 in Algorithm \ref{algo:scc}). 
    \item It is checked if the entity ID of the current IdP is already registered against \textit{CID} (the entity ID of the Combined IdP) (line 25 in Algorithm \ref{algo:scc}). If not, the entity ID is added to the list of the registered IdPs and then the list is saved in the blockchain. In addition,  a \textit{TRUE} response is sent back to the DApp (line 26 to 28 and 9 in Algorithm \ref{algo:scc}). If the current IdP is already registered, a \textit{FALSE} response is sent back to the DApp (line 30 and 9 in Algorithm \ref{algo:scc}, $M6$ in Table \ref{table:regProtocol}).
    \item The DApp then informs the admin with a meaningful message ($M7$ and $M8$ in Table \ref{table:regProtocol}).
\end{enumerate}

The admin goes through the same protocol flow to register other IdPs as well. 

\subsubsection{User registration phase} 
\begin{figure*}[h]
\includegraphics[width=0.8\linewidth]{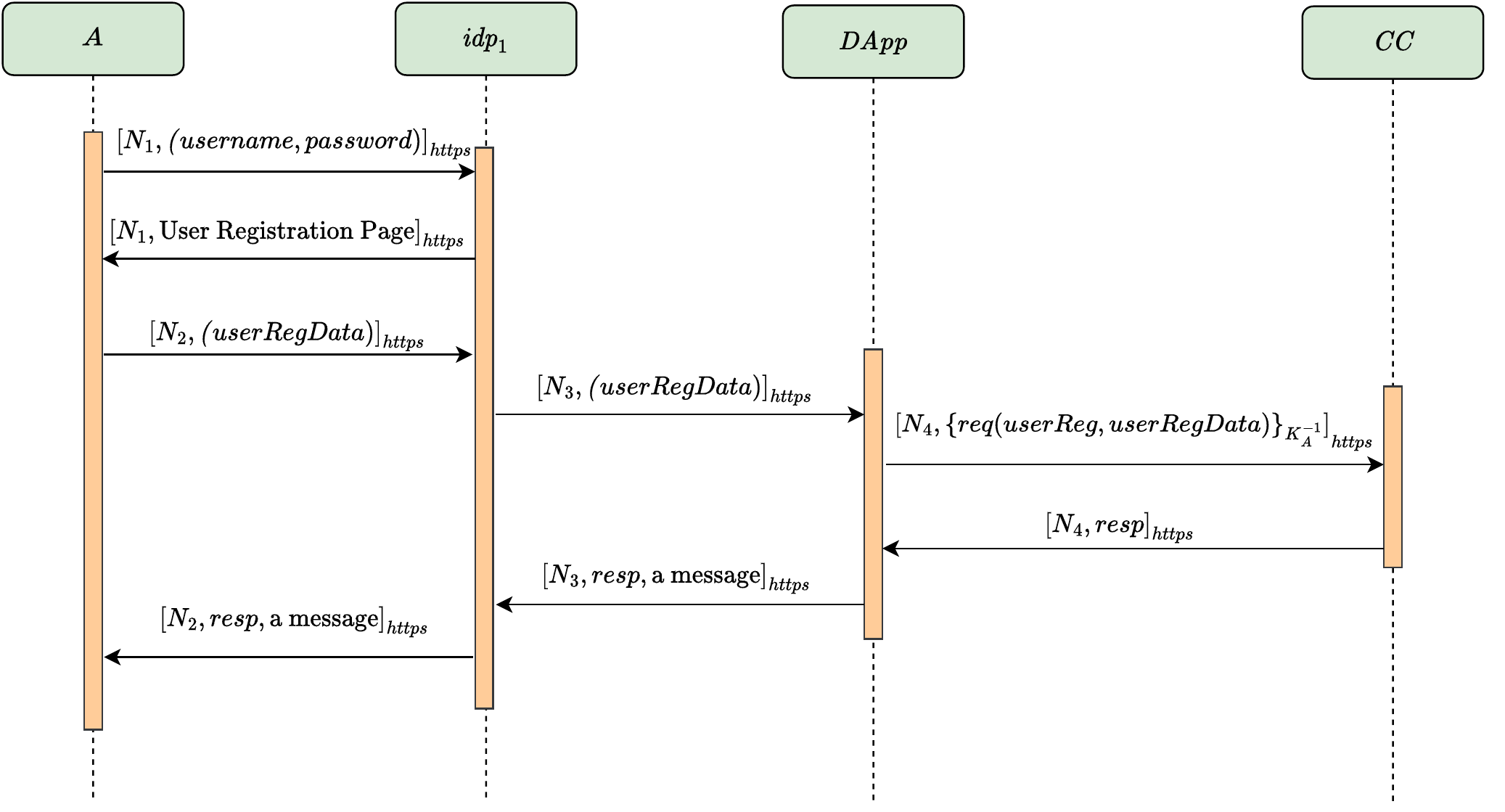}
\centering
\caption{Protocol flow for user registration}
\label{Fig:proto2}
\end{figure*}

In this phase, users are registered to one of the IdPs by the admin of the IdP. As mentioned earlier, the blockchain is used as the attribute store. When a user is registered to an IdP, their credentials and attributes are stored in the blockchain. Since all other IdPs share the same blockchain, all IdPs have access to the user credentials and attributes. It is to be noted that different IdPs may use different protocol flows for this phase. An envisioned protocol flow for this phase is illustrated in Figure \ref{Fig:proto2} and its corresponding protocol is presented in Table \ref{table:userRegProtocol}. We discuss the envisioned protocol flow next.

\begin{table}[h]
\caption{User registration protocol}
\label{table:userRegProtocol}
\centering
\begin{tabular}{p{0.2cm}lp{4.5cm}}
\hline
$M1$ & $A \rightarrow \mathit{idp}_1:$  & $[N_1, \mathit{username, password}]_{\mathit{https}}$ \\
$M2$ & $\mathit{idp}_1 \rightarrow A:$  & $[N_1, \text{User Registration Page}]_{\mathit{https}}$ \\
$M3$ & $A \rightarrow \mathit{idp}_1:$  & $[N_2, \mathit{userRegData}]_{\mathit{https}}$ \\
$M4$ & $\mathit{idp}_1 \rightarrow \mathit{DApp}:$  & $[N_3, \mathit{userRegData}]_{\mathit{https}}$ \\
$M5$ & $\mathit{DApp} \rightarrow CC:$  & $[N_4, \{\mathit{req(userReg, userRegData)\}_{K^{-1}_{A}}}]_{\mathit{https}}$ \\
$M6$ & $CC \rightarrow \mathit{DApp}:$  & $[N_4, \mathit{resp}]_{\mathit{https}}$ \\
$M7$ & $\mathit{DApp} \rightarrow \mathit{idp}_1:$  & $[N_3, \mathit{resp} \text{ with a meaningful message}]_{\mathit{https}}$ \\
$M8$ & $\mathit{idp}_1 \rightarrow A:$  & $[N_2, \mathit{resp} \text{ a meaninfgul message}]_{\mathit{https}}$ \\
\hline
\end{tabular}
\end{table}

\begin{enumerate}[i]
    \item The admin logs in to a restricted page of for the admin to $\mathit{idp}_1$ after providing their credential ($M1$ and $M2$ steps in Table \ref{table:userRegProtocol}). 
    \item After a successful login, the restricted page allows the admin to create user credentials such as a username and password for a user along with different attributes. Once completed, the admin clicks a \textit{Register} button ($M3$ in Table \ref{table:userRegProtocol}).
    \item When the \textit{Register} button is clicked, the password is hashed, and all attributes are encrypted and submitted along with other information to the corresponding DApp of the IdP ($M4$ in Table \ref{table:userRegProtocol}).
    \item The DApp creates and signs a transaction consisting of a user registration request. This request contains the type as \textit{userReg} and data as \textit{userRegData} as defined in Table \ref{table:dModel}. This request embedded within the transaction is then submitted to the blockchain to initiate the Fabric flow.
    \item Once the transaction is approved by the endorsers of the blockchain platform, it is forwarded to the invoke function of the chaincode ($M5$ in Table \ref{table:userRegProtocol}).
    \item The invoke function checks the type of the request and as it is an \textit{userReg} request, it is forwarded to the \textit{regUser} function (line 6, 7, 12 and 13 in Algorithm \ref{algo:scc}).
    \item The \textit{regUser} function retrieves the user name, hashed password, and the list of encrypted attributes from the data field (line 33 to 35 in Algorithm \ref{algo:scc}). Then, the retrieved information is stored in the blockchain and a TRUE response is returned to the DApp (line 36 and 37 in Algorithm \ref{algo:scc}, $M6$ in Table \ref{table:userRegProtocol}).
    \item The DApp then informs the admin with a meaningful message ($M7$ and $M8$ in Table \ref{table:userRegProtocol}).
\end{enumerate}

Since all the IdPs are connected to the same blockchain platform, once the user credentials and attributes are stored, all other IdPs have access to every user's data. Furthermore, all IdPs share the same symmetric encryption key so that the data can be decrypted when returning to the user. In this way, the blockchain platform is used as an attribute store. This is in contrast to the traditional SimpleSAMLPHP setup where generally a database is used.

\subsubsection{Service provisioning phase} In this phase, we discuss how the combined IdP is utilised for a SAML service provisioning activity and to simulate the notion of a decentralised identity federation. To realise this flow, we have modified */-the SP codebase of SimpleSAMLPHP so that it can interact with its corresponding DApp. Algorithm \ref{algo:dapp} presents a partial algorithm snippet for the DApp utilised in this phase.

\begin{algorithm}
\SetAlgoLined
\caption{DApp Code} \label{algo:dapp}
\SetKwBlock{Begin}{}{}
\Begin(\textbf{Start})
{
  ...\\
  send a \textit{cid} request to chaiancode and get response \textit{resp}\;
  \SetKwProg{idpResolver}{function \textbf{idpResolver}}{($\mathit{resp}$)}{}
  \idpResolver{}{
   \textit{CID} $ \leftarrow $ \textit{resp.CID}\;

    send a \textit{idpQuery} request to chaiancode and get response \textit{idps}\;
   \textit{idp} $ \leftarrow $ \textit{NULL}\;
   \For{\textit{idp} $ \in $ \textit{idps}}{
        \uIf{\textit{idp} is alive}{
            break\;
        }
    }
   send \textit{idp} back to user\;
  }
  ...\\
}
\end{algorithm}

Before we discuss the protocol flow, it is assumed that a user would like to access a service by a service provider called \textit{SP}. The SP is federated with each of the three combined IdPs following the traditional method of metadata exchange. This ensures that these IdPs trust the SP and vice versa. For simplicity, we have divided this flow into two parts: i) IdP resolving in which an IdP from the combined IdPs is selected and ii) Authentication in which the user is authenticated using the selected IdP to continue with the SAML authentication flow.

\begin{figure*}
\includegraphics[width=0.8\linewidth]{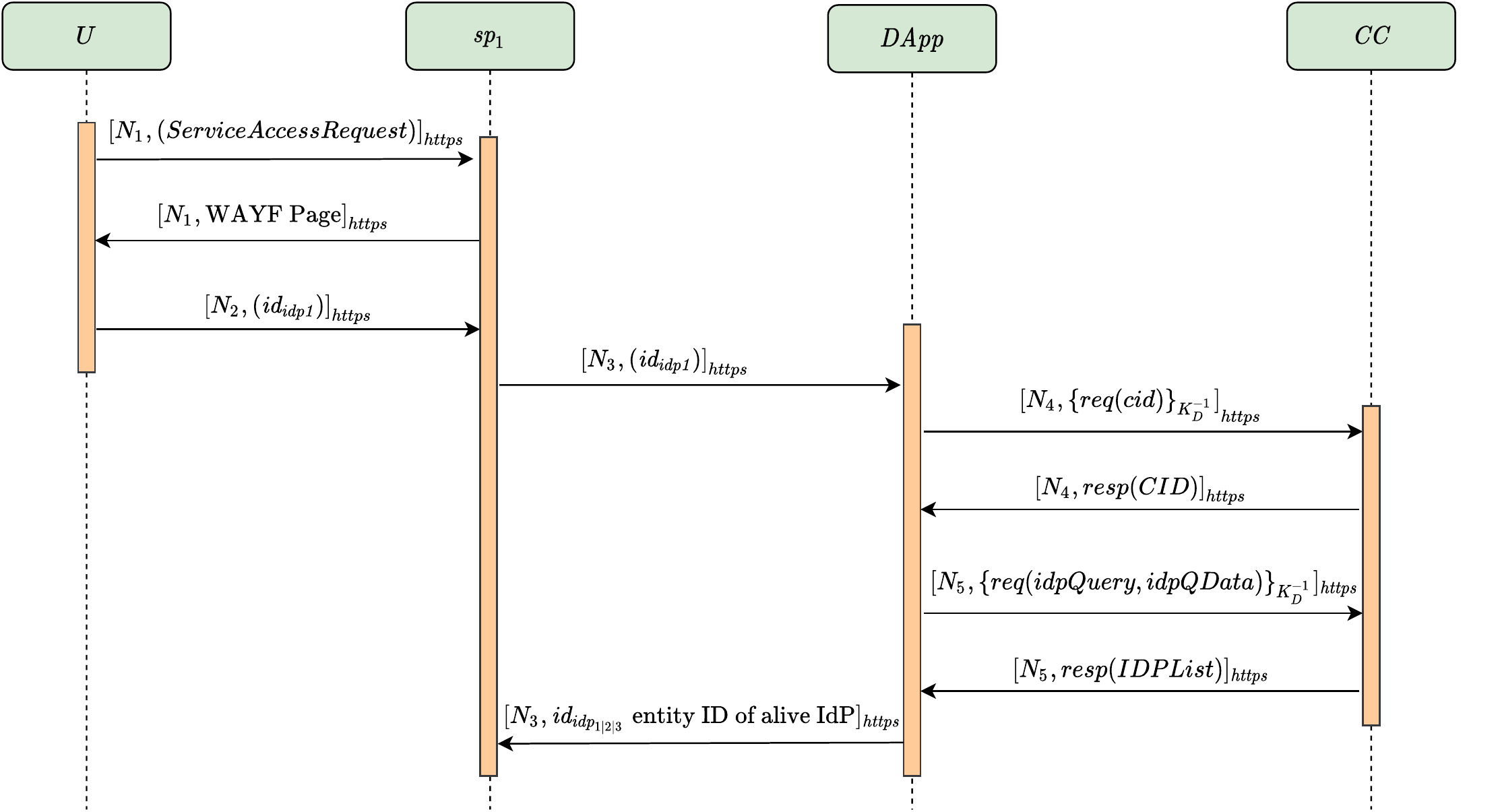}
\centering
\caption{Protocol flow for IdP resolving}
\label{Fig:proto3}
\end{figure*}

At first we present the IdP resolving flow. The flow is illustrated in Figure \ref{Fig:proto3} and its corresponding protocol is presented in Table \ref{table:idpResolvProtocol}.

\begin{table}[h]
\caption{IdP resolving protocol}
\label{table:idpResolvProtocol}
\centering
\begin{tabular}{p{0.2cm}lp{4.5cm}}
\hline
$M1$ & $U \rightarrow \mathit{sp}_1:$  & $[N_1, \text{Service access request}]_{\mathit{https}}$ \\
$M2$ & $\mathit{sp}_1 \rightarrow U:$  & $[N_1, \text{WAYF Page}]_{\mathit{https}}$ \\
$M3$ & $U \rightarrow \mathit{sp}_1:$  & $[N_2, \mathit{id}_{\mathit{idp}_1}]_{\mathit{https}}$ \\
$M4$ & $\mathit{sp}_1 \rightarrow \mathit{DApp}:$  & $[N_3, \mathit{id}_{\mathit{idp}_1}]_{\mathit{https}}$ \\
$M5$ & $\mathit{DApp} \rightarrow CC:$  & $[N_4, \{\mathit{req(cid)\}_{K^{-1}_{A}}}]_{\mathit{https}}$ \\
$M6$ & $CC \rightarrow \mathit{DApp}:$  & $[N_4, \mathit{resp(CID)}]_{\mathit{https}}$ \\
$M7$ & $\mathit{DApp} \rightarrow CC:$  & $[N_5, \{\mathit{req(idpQuery, idpQData)\}_{K^{-1}_{A}}}]_{\mathit{https}}$ \\
$M8$ & $CC \rightarrow \mathit{DApp}:$  & $[N_5, \mathit{resp(IDPList)}]_{\mathit{https}}$ \\

$M9$ & $\mathit{DApp} \rightarrow \mathit{sp}_1:$  & $[N_3, \mathit{id}_{\mathit{idp}_{1|2|3}} \text{ entity ID of alive IdP}]_{\mathit{https}}$ \\
\hline
\end{tabular}
\end{table}

\begin{enumerate}[i]
    \item A user visits the service page of the SP for the first time ($M1$ in Table \ref{table:idpResolvProtocol}).
    \item The user is forwarded to the SAML WAYF (Where Are You From) page of the SP where the user can select an IdP (($M2$ in Table \ref{table:idpResolvProtocol})).
    \item The WAYF pages shows the entity IDs of the three IdPs separately. The user selects one of the IdPs ($M3$ in Table \ref{table:idpResolvProtocol}).
    \item Internally, the entity ID of selected IdP is sent to the DApp ($M4$ in Table \ref{table:idpResolvProtocol}).
    \item The DApp creates and signs a transaction consisting of a \textit{cid} request which does not have any data field. This request embedded within the transaction is then submitted to the blockchain to initiate the Fabric flow.
    \item Once the transaction is approved by the endorsers of the blockchain platform, it is forwarded to the invoke function of the chaincode ($M5$ in Table \ref{table:idpResolvProtocol}).
    \item The invoke function checks the type of the request and as it is a \textit{cid} request, the $\mathit{CID}$ value is returned as a response. (line 16 to 18 in Algorithm \ref{algo:scc}, $M6$ in Table \ref{table:idpResolvProtocol}).
    \item Then, the DApp calls the \textit{idpResolver} function with the response \textit{resp} as a parameter (Line 4 in Algorithm \ref{algo:dapp}).
    \item Within the function, the CID is retrieved from the response \textit{resp} (Line 6 in Algorithm \ref{algo:dapp}).
    \item The DApp creates and signs another transaction consisting of a \textit{idpQuery} request which contains idqQData as data (as modeled in Table \ref{table:dModel}). This request embedded within the transaction is then submitted to the blockchain to initiate the Fabric flow.
    \item Once the transaction is approved by the endorsers of the blockchain platform, it is forwarded to the invoke function of the chaincode ($M7$ in Table \ref{table:idpResolvProtocol}).
    \item The invoke function checks the type of the request and as it is an \textit{idpQuery} request, it is forwarded to the \textit{queryIDP} function (line 6, 7, 10 and 11 in Algorithm \ref{algo:scc}).
    \item Within the \textit{queryIDP} function, CID is used to retrieve the list of the entity IDs of the three combined IdPs and return the list to the DApp (line 20, 21 and 11 in Algorithm \ref{algo:scc}). 
    \item Using the CID, the list of the entity IDs of the combined IdPs is returned using the \textit{getIDPList} function of the chaincode (Line 7 in Algorithm \ref{algo:dapp}, $M8$ in Table \ref{table:idpResolvProtocol}).
    \item Then, the DApp loops through the list of entity IDs to check if any of the IdPs is alive. For this, the DApp just checks if it can retrieve the metadata of the IdP from the entity ID URL. The entity ID of the first IdP that is found to be alive is returned to the WAYF Page (Line 8 to 12 in Algorithm \ref{algo:dapp}, $M9$ in Table \ref{table:idpResolvProtocol}).
\end{enumerate}

\begin{figure*}
\includegraphics[width=1\linewidth]{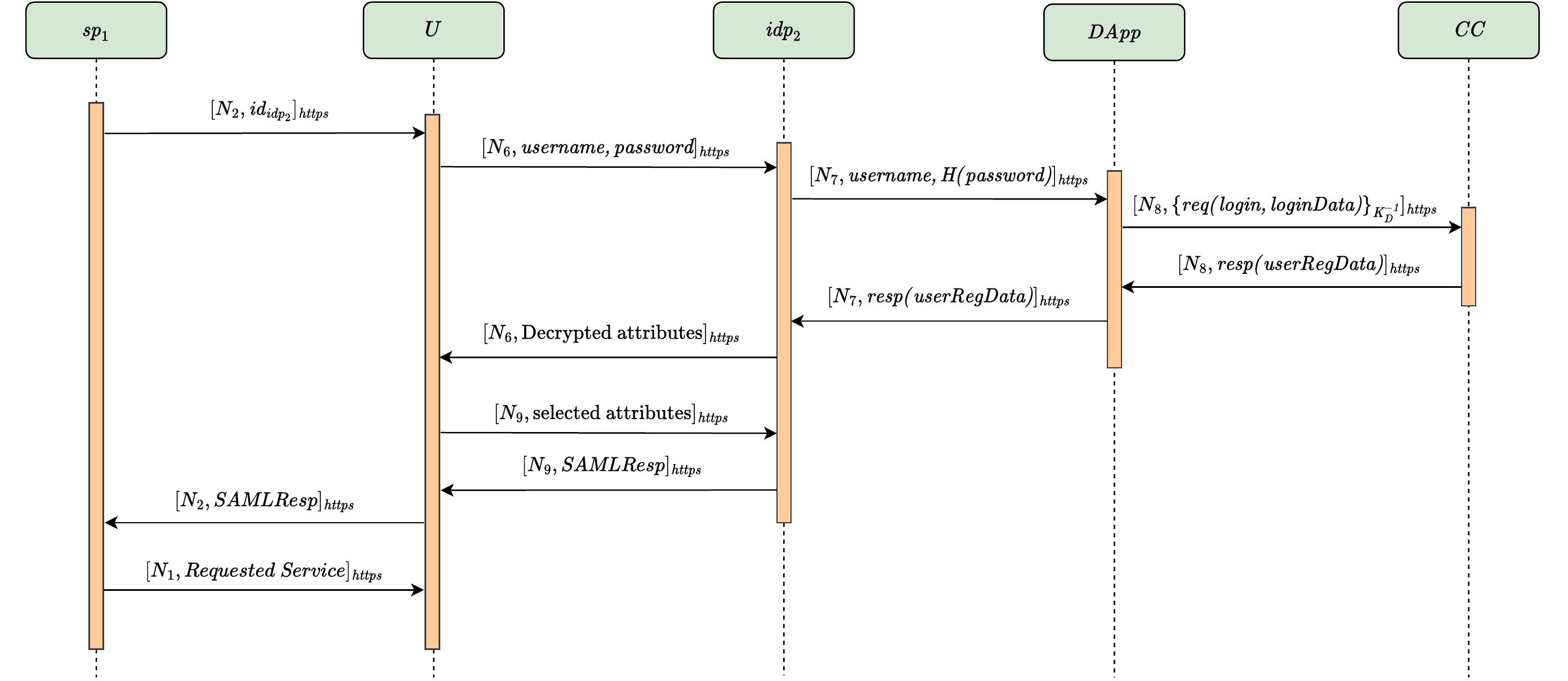}
\centering
\caption{Protocol flow for login}
\label{Fig:proto4}
\end{figure*}

Next, we present the SAML authentication flow. The flow is illustrated in Figure \ref{Fig:proto4} and its corresponding protocol is presented in Table \ref{table:loginProtocol}. 

\begin{table}[h]
\caption{Login protocol}
\label{table:loginProtocol}
\centering
\begin{tabular}{p{0.5cm}lp{4.5cm}}
\hline
$M10$ & $\mathit{sp}_1 \rightarrow U:$  & $[N_2, \mathit{id}_{\mathit{idp}_{2}}]_{\mathit{https}}$ \\
$M11$ & $U \rightarrow \mathit{idp}_2:$  & $[N_6, \mathit{username, password}]_{\mathit{https}}$ \\
$M12$ & $\mathit{idp}_2 \rightarrow \mathit{DApp}:$  & $[N_7, \mathit{username, H(password)}]_{\mathit{https}}$ \\
$M13$ & $\mathit{DApp} \rightarrow CC:$  & $[N_8, \{\mathit{req(login, loginData)\}_{K^{-1}_{A}}}]_{\mathit{https}}$ \\
$M14$ & $CC \rightarrow \mathit{DApp}:$  & $[N_8, \mathit{resp(\text{encrypted attributes})}]_{\mathit{https}}$ \\
$M15$ & $\mathit{DApp} \rightarrow \mathit{idp}_2:$  & $[N_7, \mathit{resp(\text{encrypted attributes})}]_{\mathit{https}}$ \\
$M16$ & $\mathit{idp}_2 \rightarrow U:$  & $[N_6, \text{Decrypted attributes}]_{\mathit{https}}$ \\
$M17$ & $U \rightarrow \mathit{idp}_2:$  & $[N_9, \text{selected attributes}]_{\mathit{https}}$ \\
$M18$ & $\mathit{idp}_2 \rightarrow U:$  & $[N_9, \textit{SAMLResp}]_{\mathit{https}}$ \\
$M19$ & $U \rightarrow \mathit{sp}_1:$  & $[N_1, \textit{SAMLResp}]_{\mathit{https}}$ \\
$M20$ & $\mathit{sp}_1 \rightarrow U:$  & $[N_1, \text{Requested service}]_{\mathit{https}}$ \\
\hline
\end{tabular}
\end{table}

\begin{enumerate}[i]    
    \item The SAML SP codebase forwards the user to the selected IdP (assumed $\mathit{idp}_2$ to be alive in the protocol flow in the $M10$ step in Table \ref{table:loginProtocol}) with a SAML Authentication request (a request with its type as \textit{authn} and data as \textit{AuthnReq} as presented in Table \ref{table:dModel}) where the user needs to authenticate.
    \item For authentication, the IdP shows the user a login page where the user submits their username and password and clicks the \textit{Submit} button ($M11$ in Table \ref{table:loginProtocol}). 
    \item The IdP hashes the password and submits the username and hashed passport to the DApp ($M12$ in Table \ref{table:loginProtocol}).
    \item DApp creates a login request consisting of \textit{login} as the type and \textit{logInData} as the data as per Table \ref{table:dModel}. Then, the DApp submits a transaction consisting of the request to the blockchain to initiate the Fabric flow.
    \item Once the transaction is approved by the endorsers of the blockchain platform, it is forwarded to the invoke function of the chaincode ($M13$ in Table \ref{table:loginProtocol}).
    \item The invoke function checks the type of the request and as it is a \textit{login} request, it is forwarded to the \textit{loginUser} function (line 6, 14 and 15 in Algorithm \ref{algo:scc}).
    \item Within the login function, the username and password hash are retrieved from the request (line 39 in Algorithm \ref{algo:scc}) and then uses the username to retrieve stored hashes from the blockchain (line 40 \& 41 in Algorithm \ref{algo:scc}).
    \item Then two hashes are matched. If they match, stored attributes of the users are returned to the DApp (line 44 \& 16 in Algorithm \ref{algo:scc}, $M14$ in Table \ref{table:loginProtocol}). If they do not match, a FALSE response is returned to the DApp (line 46 \& 16 in Algorithm \ref{algo:scc})
    \item The response is returned to the IdP ($M15$ in Table \ref{table:loginProtocol}). Depending on the response, either user attributes are decrypted and are shown to the user ($M16$ in Table \ref{table:loginProtocol}) or a meaningful message (e.g. \textit{username/password do not match}) is shown to the user.
    \item Assuming a positive response, the user can choose the attributes that they would like to release to the SP and once selected, the user clicks the \textit{Submit} button ($M17$ in Table \ref{table:loginProtocol}).
    \item The IdP creates a SAML assertion, signs the assertion and embeds into a SAML response. The SAML response (resp as modelled in Table \ref{table:dModel}) is returned to the SP via the user ($M18$ and $M19$ steps in Table \ref{table:loginProtocol}).
    \item The SP gets the assertion from the response, validates its signature and retrieves the user attributes. Based on the attributes, the SP takes an authorisation decision to allow (or reject) the user to access the service ($M20$ in Table \ref{table:loginProtocol}).
\end{enumerate}

The cornerstone of this approach is that the SP will be able to offer federated services as long as there is at least one IdP is available when IdPs become unavailable (e.g. due to a technical reasons or an attack). This has been possible because of the utilisation of blockchain to create the notion of a combined IdP which shares the same credential and attribute data storage facility within the blockchain. In this way, we have been able decentralise an identity federation and remove the bottleneck of a single point of failure because of its reliance on a single IdP. The best part is that the user does need to worry at all about this process, the intricate complex protocol flows are tackled under the hood and the user experience will be exactly similar to the existing flow.

\section{Evaluation}
\label{sec:analysis}

In this section, we assess and analyse the performance of a traditional identity federation using SimpleSAMLphp and compare it to the performance of the Decentralised Identity Federation (DIF) that we proposed and developed. In this experiment, we used throughput, latency, and resource consumption as performance metrics. With these metrics, we conducted several load-testing experiments for multiple scenarios, modeling many use cases and configurations. The use cases involved regular SAML interactions between users, IdPs, and SPs as explained next.

\begin{itemize}
    \item \textbf{Registering Users:} In this use case, we create a group of user accounts and store their credentials within the Identity Provider (IdP) to be used for authentication later.
    
    \item \textbf{Requesting Service:} A user first submits a request to an SP for accessing a specific service. In our test scenarios, 2 SPs offer 2 separate services. To utilise a service from any SP, the user must first be authenticated by the federation's IdP.
    
    \item \textbf{Authenticating Users:} The user is subsequently redirected to the selected IdP's authentication page via a SAML authentication request. When the IdP receives the request, it checks to see if the SP is a federation member. Then the user authenticates at the IdP and receives a SAML response containing a SAML assertion which is forwarded back to the SP.
    
    \item \textbf{Responding to Service Request:} The assertion is retrieved from the SAML response when the SP receives it. The assertion's signature is then verified, and the user's attributes are extracted from the assertion. The user is then allowed to access the resource.

\end{itemize}

We used two separate test plans to investigate the use cases during our study. We addressed the user registration process in the first test plan and simulated a typical user interaction consisting of service request, authentication, and response in the second test plan.

\subsection{Experimental Setup}

\begin{figure}
\includegraphics[width=1\linewidth]{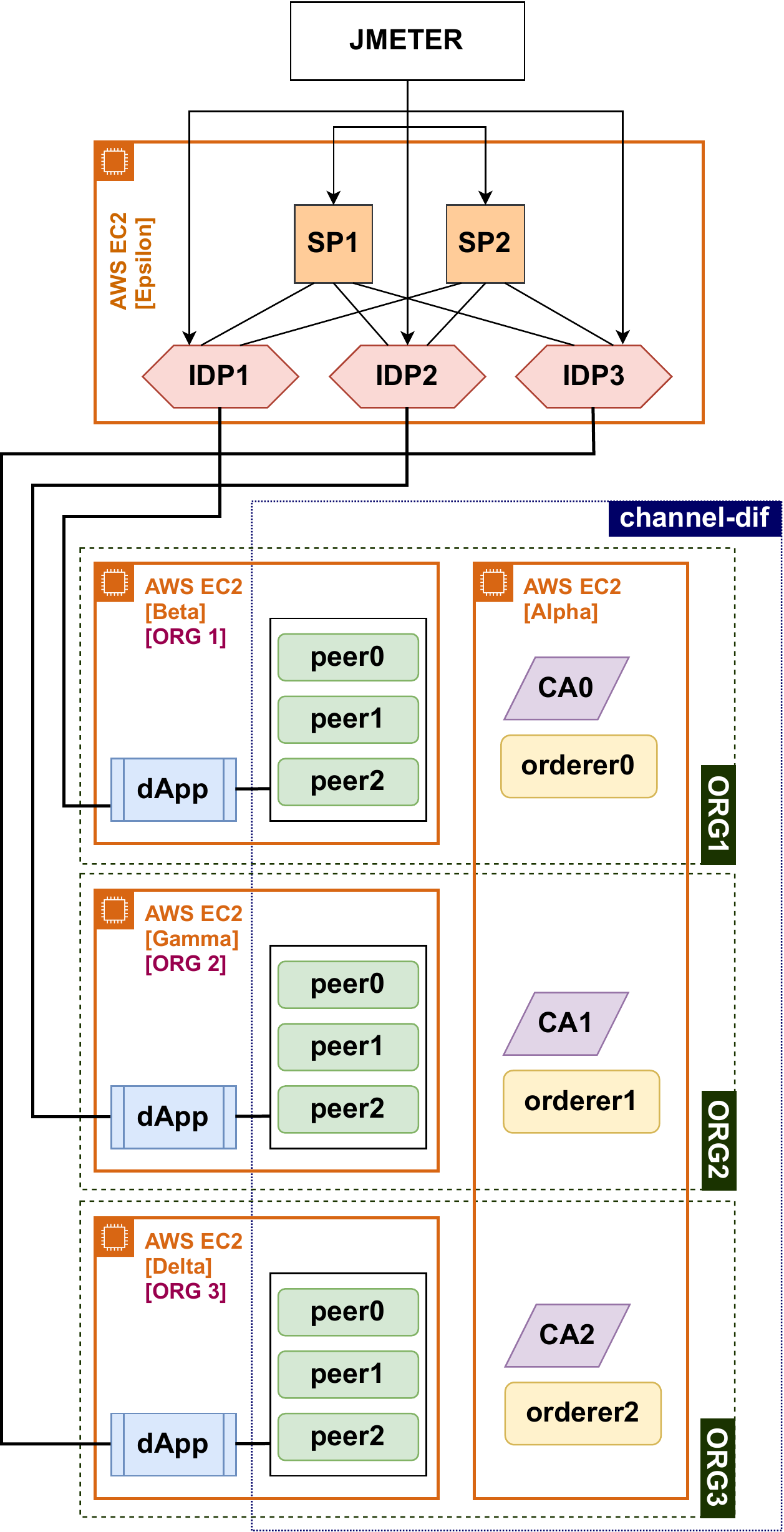}
\centering
\caption{Network configurations for experiments}
\label{Fig:DIF_experimental_setup}
\end{figure}

For our experimental evaluation, we had 3 IdPs and 2 SPs. When using the Decentralised Identity Federation, each IdP participated in the blockchain network as a single organisation. Our system was deployed in Amazon Web Services (AWS) EC2 instances. The whole network was built using 5 Ubuntu instances, each with 4 VCPUs and 16 GBs of RAM. The network configuration is shown in Figure \ref{Fig:DIF_experimental_setup}.

Here, 3 instances (Beta, Gamma, Delta) were used to represent the 3 participating IdP organisations (ORG1, ORG2, ORG3) and they hosted the necessary docker containers. The Orderer nodes and Certificate Authorities were hosted on the 4th EC2 instance (Alpha), while the SPs and IdPs were hosted on the 5th (Epsilon). Each organisation has 3 peers (peer0, peer1, peer2) and a dApp running to facilitate interaction with the network. The certificate authority of each organisation was hosted separately in the Alpha instance, and the ordering service consisted of 2 or 3 orderers (based on an experimental setup) with 3 ZooKeeper and 4 Kafka brokers.

To execute the test plans, we used Apache JMeter \cite{jmeter}, an open-source performance measurement and load testing software. We carried out the load tests by gradually increasing the traffic. We started with 10 simultaneous users, and increased the number of users after each iteration of the test plan, going up to 150 simultaneous users. Each test plan was executed against 3 different experimental setups to investigate their impact on performance, 1) Conventional SAML setup without DIF (no-blockchain), 2) SAML with DIF using Hyperledger Fabric (2 orderers), 3) SAML with DIF using Hyperledger Fabric (3 orderers).

In the conventional SAML setup (i.e. no-blockchain setup), we used MySql \cite{mysql} database for storing the credentials. However, in the other two blockchain-based network setups we took advantage of levelDB \cite{leveldb} for storing the world state of the blockchain.

The CPU and RAM consumption in all of the test situations stays within a fair range and never approaches the system limit. Similarly, increasing the number of orderers does not have much affect on the CPU usage, but RAM usage increases slightly when we go from 2 orderers to 3 orderers.

\vspace{-3mm}
\begin{figure*}[ht]
  \begin{subfigure}{0.49\textwidth}
    \includegraphics[width=\linewidth]{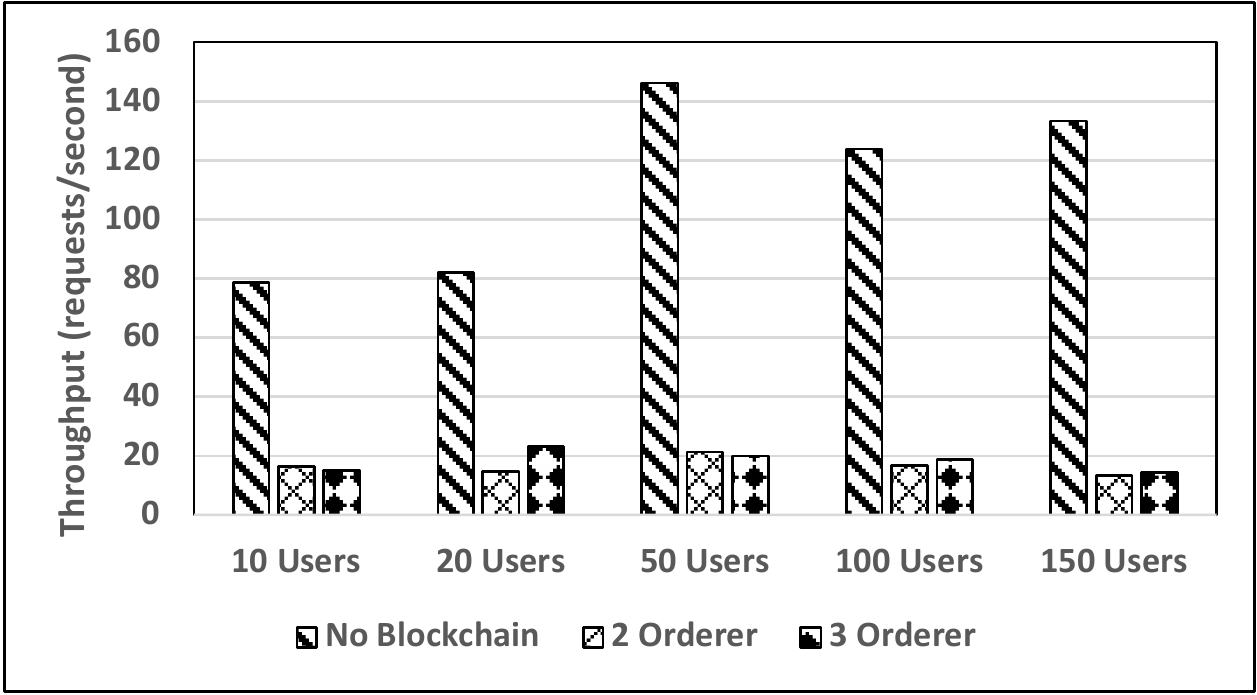}
    \caption{Throughput vs Load}  \label{fig:throughput_signup}
  \end{subfigure}
  \hspace*{\fill} 
  \begin{subfigure}{0.49\textwidth}
        \includegraphics[width=\linewidth]{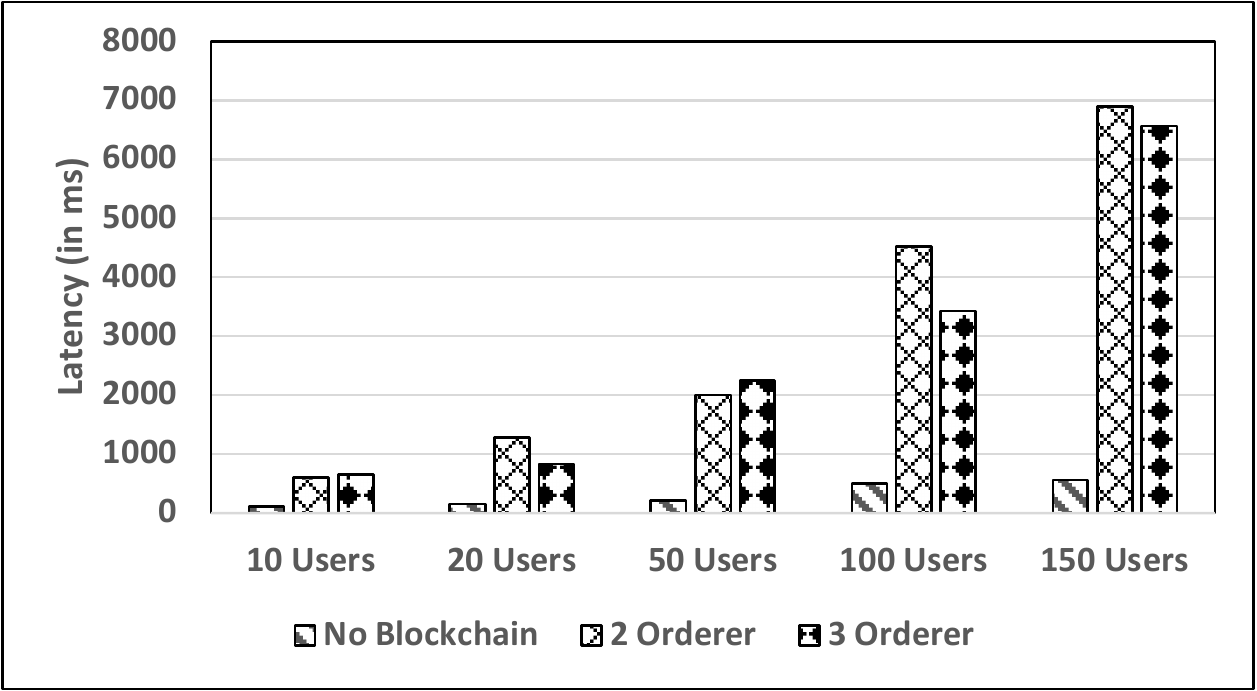}
    \caption{Latency vs Load}  \label{fig:latency_signup}
  \end{subfigure}
  \caption{Registration Performance evaluation (Throughput and Latency)} \label{fig:throughput_latency_signup}

\vspace{-1mm}  
  \begin{subfigure}{0.49\textwidth}
        \includegraphics[width=\linewidth]{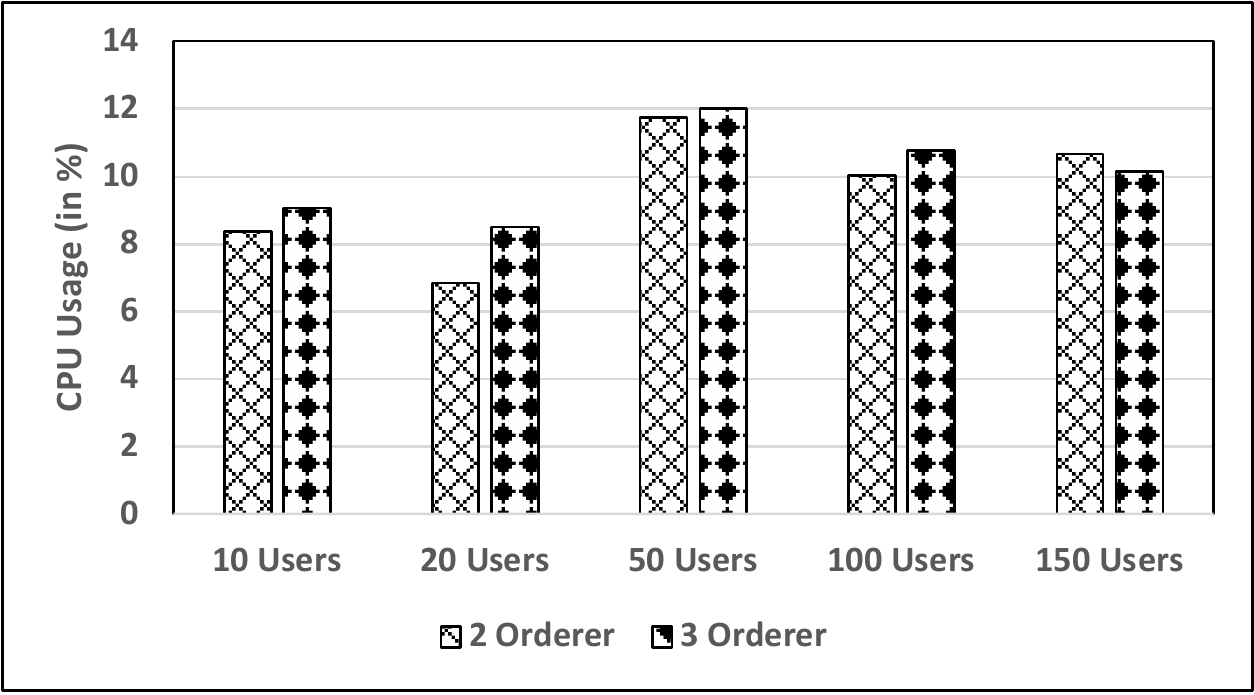}
    \caption{CPU Usage vs Load}  \label{fig:cpu_usage_signup}
  \end{subfigure}
  \hspace*{\fill} 
  \begin{subfigure}{0.49\textwidth}
        \includegraphics[width=\linewidth]{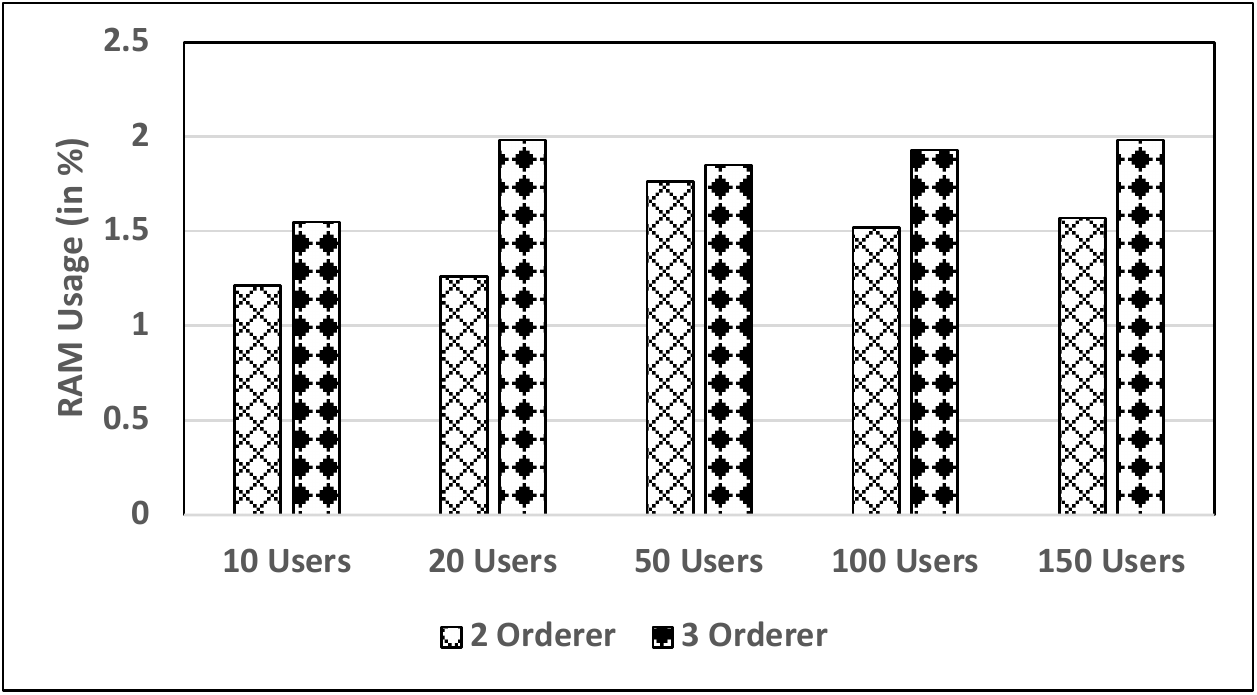}
    \caption{RAM Usage vs Load}  \label{fig:ram_usage_signup}
  \end{subfigure}
  \caption{Registration Resource Consumption (CPU and RAM)} \label{fig:resource_usage_signup}

\vspace{-1mm}  
  \begin{subfigure}{0.49\textwidth}
        \includegraphics[width=\linewidth]{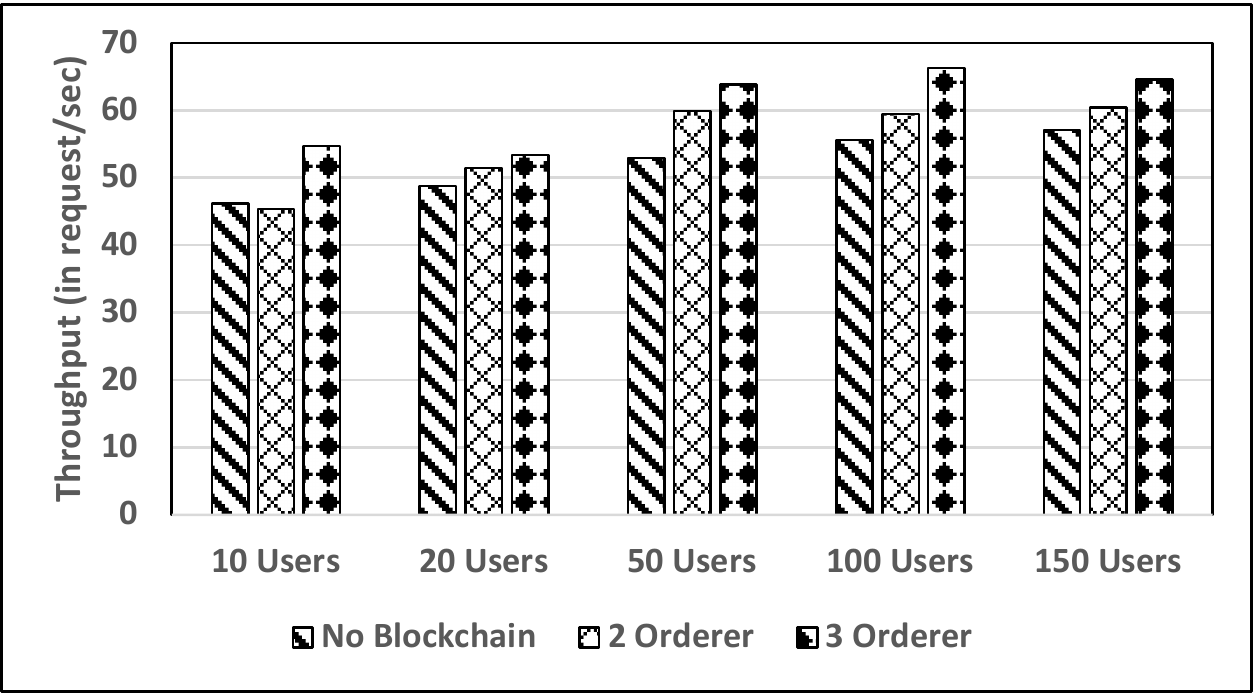}
    \caption{Throughput vs Load}  \label{fig:throughput_login}
  \end{subfigure}
  \hspace*{\fill} 
  \begin{subfigure}{0.49\textwidth}
        \includegraphics[width=\linewidth]{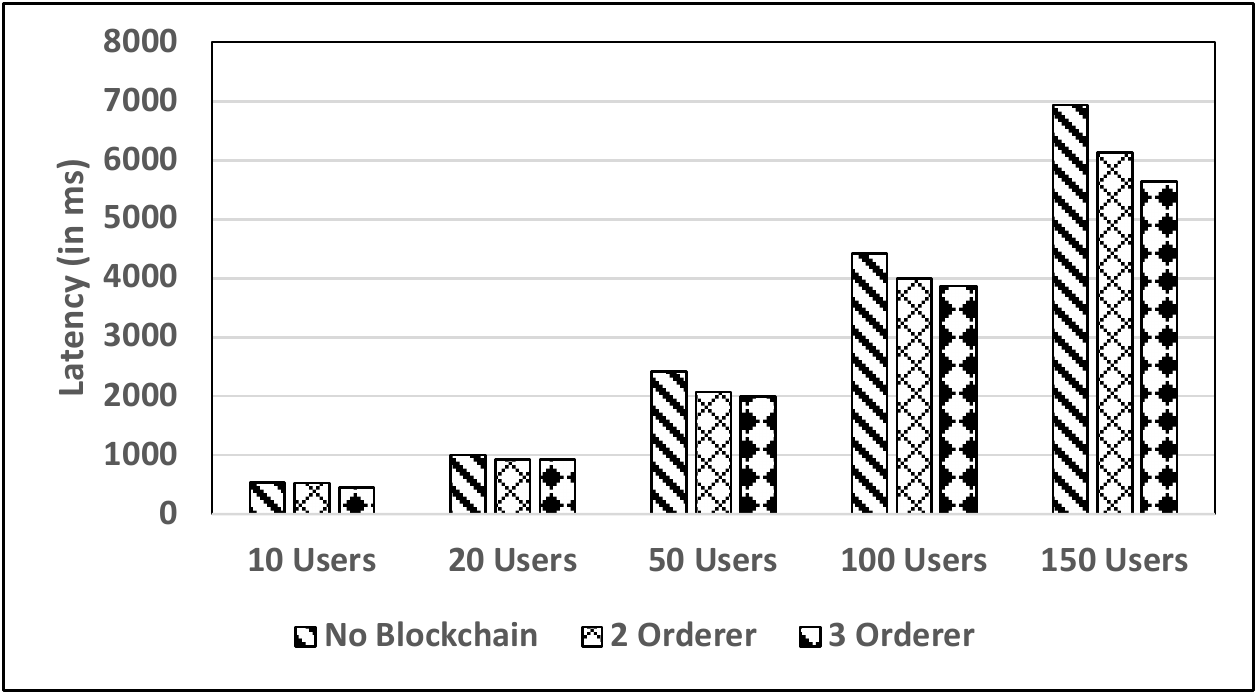}
    \caption{Latency vs Load}  \label{fig:latency_login}
  \end{subfigure}
  \caption{Login Performance evaluation (Throughput and Latency)} \label{fig:throughput_latency_login} 

\vspace{-1mm} 
  \begin{subfigure}{0.49\textwidth}
        \includegraphics[width=\linewidth]{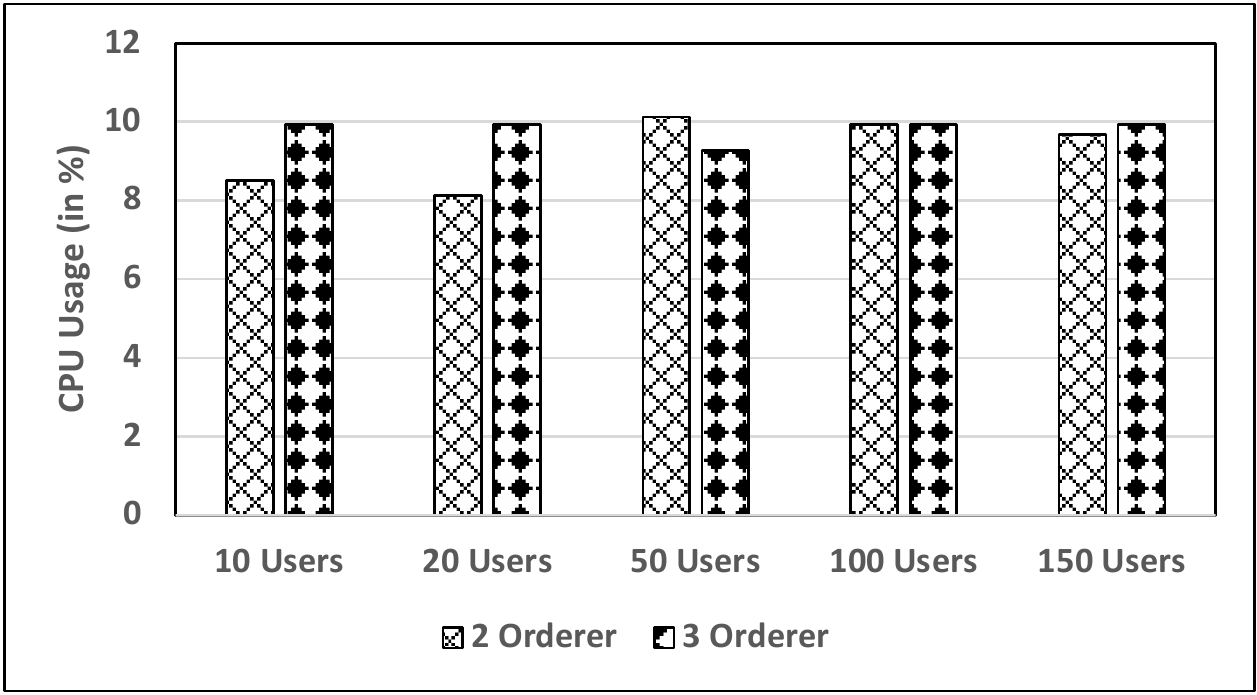}
    \caption{CPU Usage vs Load}  \label{fig:cpu_usage_login}
  \end{subfigure}
  \hspace*{\fill} 
  \begin{subfigure}{0.49\textwidth}
        \includegraphics[width=\linewidth]{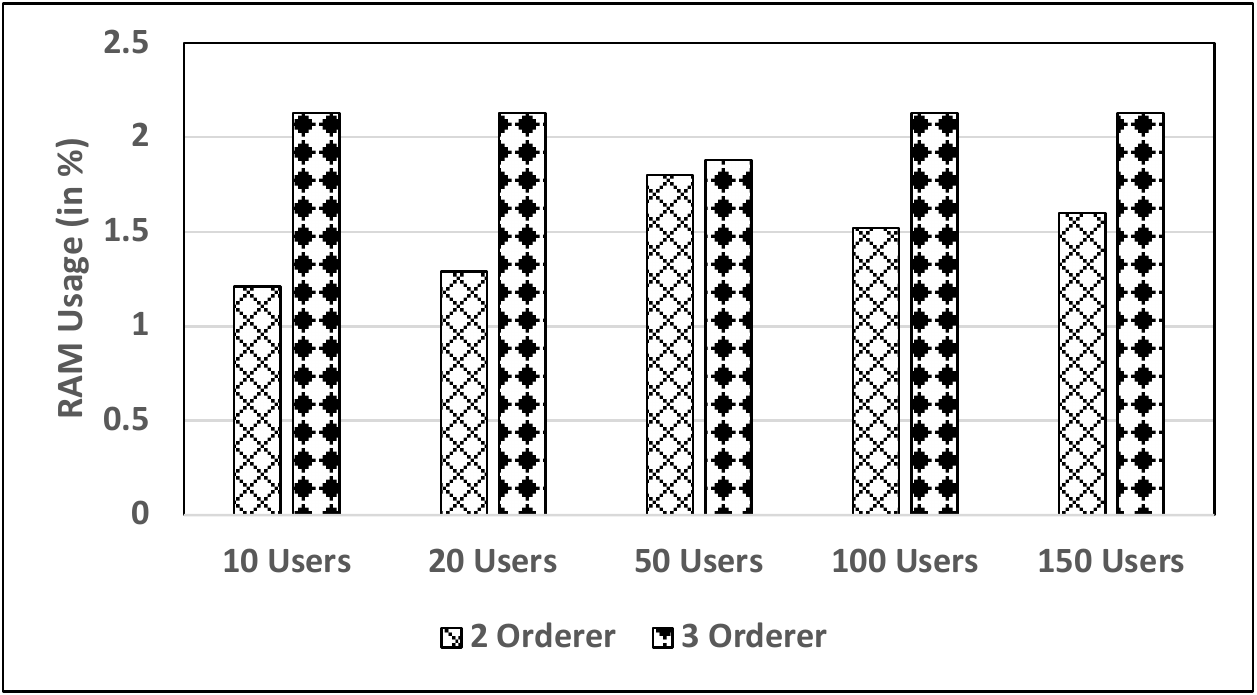}
    \caption{RAM Usage vs Load}  \label{fig:ram_usage_login}
  \end{subfigure}
  \caption{Login Resource Consumption (CPU and RAM)} \label{fig:resource_usage_login}
\vspace{-3mm}  

\end{figure*}

\subsection{Performance Analysis}
We recorded the throughput, latency and resource consumption (memory and CPU  usage) after each test cycle for comparing with the subsequent cycles. After completing the test, data from different setups and varying loads were aggregated to investigate the performance under different circumstances. Figure \ref{fig:throughput_latency_signup}, Figure \ref{fig:cpu_usage_signup}, Figure \ref{fig:ram_usage_signup}, Figure \ref{fig:throughput_latency_login}, Figure \ref{fig:cpu_usage_login} and Figure \ref{fig:ram_usage_login} show the comparative evaluation results of the two test plans (registration and login usage) in different situations.

\subsubsection{Test Plan}
\begin{itemize}
    \item \textit{Registration latency and throughput}: Figure \ref{fig:throughput_latency_signup} depicts that, the no-blockchain setup of the network has much better throughput and lower latency than the other two blockchain network configurations. We can also see that the throughput in the No-Blockchain SAML setup increases with the load, which indicates that it has not reached the maximum capacity yet. However, in the blockchain setups, the throughput doesn't change much because complex communication within the network limits the maximum throughput quite early.

    \item \textit{Registration resource consumption}: From Figure \ref{fig:resource_usage_signup}, the resource consumption, in terms of both CPU and RAM usages, for all load plans are considerably low. However, there are two trends: i) resource consumption for both resources tend to increase as more user loads are added and ii) resources consumption is higher for 3 orderers than 2 orderers. These trends are expected as more users and more orderers would imply more computations and storage are required.
    
    \item \textit{Login latency and throughput}: From Figure \ref{fig:throughput_latency_login}, we can see that, latency of the system increases linearly as more loads are added. Similarly, the throughput also increases with the load. However, the change in throughput is not as steep as the latency. This is because once the throughput reaches its max capacity, after a certain amount of load, the throughput does not change as much even if it has to handle more loads. Using LevelDB helps our system execute queries faster than a relational database (MySQL) in a no-blockchain setup.

    An important observation from Figure \ref{fig:throughput_latency_signup} Figure \ref{fig:throughput_latency_login} is that the performance of the system varies significantly between the test plans. In the Registration test plan (Figure \ref{fig:throughput_latency_signup}), typical SAML implementation (no blockchain) outperforms our Decentralised Identity Federation. However, during the login process (Figure \ref{fig:throughput_latency_login}), our system has slightly better performance than the typical SAML. This outcome is expected because, during the Registration phase, all the organisations must reach a consensus for creating the block that updates the world state, whereas, during the login process, the peers make queries on the local ledger which does not have the consensus reaching overhead. In a practical scenario, a user will go through the Registration process only once, but the login phase will be executed many more times whenever the user needs to authenticate to the system. Consequently, implementation of the proposed system will not hinder performance in most usage scenarios.


    \item \textit{Login resource consumption}: It can be observed from from Figure \ref{fig:resource_usage_login} that the CPU and RAM usages for all load plans in the login process are considerably low. Also, like before, both CPU and RAM usages tend to increase as more user loads are added and resources consumption is higher for 3 orderers than 2 orderers.   
\end{itemize}

\section{Discussion}
\label{sec:discussion}
Throughout this article, we have presented different aspects of our proposal of decentralising the identity provider entity of any existing Identity Federation through the use of blockchain. We have explained the motivation, outlined the proposal, presented the architecture and the implementation details along with protocol flows and finally, analysed its performance. In this section, we explain how the architecture and its implementation have satisfied different functional and security requirements (Section \ref{subsec:reqAnalysis}) and analyse its security (Section \ref{subsec:sec}). Furthermore, we also discuss its advantages and potential future work in Section \ref{subsec:advantages}. 

\subsection{Analysing Requirements}
\label{subsec:reqAnalysis}
Through the designing and implementation of our system we were able to meet all of the requirements for the system, both functional and non-functional. The combined IdPs are part of the same private blockchain network which binds their trust with each other. This is implicitly analogous to the trust establishment procedure within  federated entities by exchanging metadata and hence, it satisfies requirement \textit{F1}. These combined IdPs utilise the blockchain as the metadata store, thereby sharing the user attributes with each other to satisfy requirement \textit{F2}. The protocol ensures that even when one of the IdPs from the combined IdPs is offline, SPs can leverage the service of other IdPs in a seamless way. The blockchain based attribute storage facility ensures that any of the IdPs have access to the user attributes and carry out the SAML authentication functionality. This satisfies requirement \textit{F3}. Finally, the protocol is designed in such a way that SAML SPs are minimally affected and hence, it satisfies requirement \textit{F4}.

Each SP is federated with each of the combined IdPs in the traditional approach by exchanging SAML metadata manually. This is usually carried out by the admins of the respective entity and hence, it satisfies requirement \textit{S1}. Each assertion is digitally signed to satisfy requirement \textit{S2}. An IdP releases only encrypted assertions and all data are transmitted via HTTPS channels, thereby satisfying requirement \textit{S3}. The consent module in the SimpleSAMLPHP ensures that user attributes are released only after an explicit consent by a user, this satisfies \textit{S4}. Finally, even if an IdP from the combined IdP becomes unavailable, the protocol of our system ensures the availability of IdP services within the federation. This satisfies \textit{S4}.

\subsection{Protocol Verification}
\label{subsec:sec}
As discussed in the previous section, the proposed architecture and the protocol have satisfied different security requirements. Even so, the architecture and its implementation employ complex interactions with different components. It is well understood that the security of any system having complex interactions is difficult to ensure. One prominent approach to analyse the security of any such system is to formally analyse its security. Towards this aim, we have formally analysed the security of our approach using a state-of-the-art Protocol Verification tool called ProVerif \cite{BlanchetFOSAD14, BlanchetSmythJCS18}. Using ProVerif, we will analyse if the protocol of the developed system satisfies the secrecy as well as authentication goals. The secrecy goals checks if  a secret value truly remains secret while being transmitter between  two entities \cite{ferdous2016formalising}. On other hand, the authentication goals checks the authenticity of two entities while data are being transmitted between them.  In the following, we discuss how we have formalised our protocol using ProVerif to evaluate the secrecy and authentication goals of the protocol. 

\subsubsection{Secrecy Goals}
\label{subsubsec:sec_goals}

In our ProVerif script, we declare some private variables. These variable represent sensitive information that we are trying to communicate between different entities. A brief introduction of these variables are described below.

\begin{itemize}
    \item \textit{username, password}: The username and password of the admin/user.
    \item \textit{IdpRegPage, IdpRegData}: The $idp$ registration page and data.
    \item \textit{UserRegPage, UserRegData}: The $user$ registration page and data.
    \item \textit{IdpAlive}: The identity of an $idp$ that is alive.
    \item \textit{samlResp}: The $SAML$ response data.
    \item \textit{decAtt}: The decrypted attributes.
    \item \textit{CID}: Common Entity ID for the combined $CoT$
\end{itemize}

To analyse the reachability of these variables, we write the following lines of code to encode our secrecy queries for the protocols described in Table \ref{table:regProtocol}, Table  \ref{table:userRegProtocol}, Table  \ref{table:idpResolvProtocol} and Table  \ref{table:loginProtocol}.

\begin{lstlisting}[caption={Secrecy Queries},captionpos=b,frame=tb]
(* User registration protocol *)
query attacker(username);
      attacker(password);
      attacker(UserRegData);
      attacker(UserRegPage).

(* Idp registration protocol *)
query attacker(username);
      attacker(password);
      attacker(IdpRegPage);
      attacker(IdpRegData).

(* Idp registration and Login protocol *)
query attacker(IdpAlive);
      attacker(password);
      attacker(username);
      attacker(samlResp);
      attacker(decAtt);
      attacker(CID).
\end{lstlisting}

\subsubsection{Authenticity Goals}
\label{subsubsec:auth_goals}
Correspondence assertions are used in ProVerif  to analyse the link between occurrences. The protocol's validity is tested by this behaviour \cite{woo1993semantic}. We highlight key stages in our protocols described in Table \ref{table:regProtocol}, Table  \ref{table:userRegProtocol}, Table  \ref{table:idpResolvProtocol} and Table  \ref{table:loginProtocol} with the following events:

\begin{itemize}
    \item \textbf{event} $beginLoginReq$ and $endLoginReq$ mark the starting and termination points of the login request event.
    \item \textbf{event} $beginServReqVal$  ($endServReqVal$) marks the start (resp. the end) of the validation that happens at the $idp$ end to determine which service the requesting entity has access to.
    \item \textbf{event} $beginRegPageReq$ and $endRegPageReq$ are triggered when an admin/user requests the registration page or the registration is served respectively.
    \item \textbf{event} $beginIdpReg$ ($endIdpReg$) marks the beginning (resp. the end) of registering an entity in the identity provider, $idp$.
    \item \textbf{event} $beginSndIdpReg$ ($endSndIdpReg$) takes place when the $DApp$ sends (resp. receives) $idp$ registration data to (resp. from) $CC$.
    \item \textbf{event} $beginUserReg$ ($endUserReg$) triggers the beginning (resp. ending) of registering a user.
    \item \textbf{event} $beginSndUsrReg$ ($endSndUsrReg$) takes place when the $DApp$ sends (resp. receives) admin registration data to (from) $CC$.
\end{itemize}

After defining the events that make up the core of our protocol, we can now create correspondence queries for these events to verify our authenticity objectives. The below list contains the correspondences.

\begin{lstlisting}[caption={Idp registration protocol},captionpos=b,frame=tb]
(* Idp registration protocol *)
query admin:host, idp1:host, nonce:nonce; 
    event(endRegPageReq(admin,idp1,nonce)) 
    ==> 
    event(beginRegPageReq(admin,idp1,
    nonce)).
    
query admin:host, idp1:host, nonce:nonce; 
    event(endIdpReg(admin,idp1,nonce)) 
        ==> 
    event(beginIdpReg(admin,idp1,nonce)).
    
query dapp:host, cc:host, nonce:nonce; 
    event(endSndIdpReg(dapp,cc,nonce)) 
        ==> 
    event(beginSndIdpReg(dapp,cc,nonce)).
\end{lstlisting}

\begin{lstlisting}[caption={User registration protocol},captionpos=b, frame=tb]
(* User registration protocol *)
query admin:host, idp1:host, nonce:nonce; 
    event(endRegPageReq(admin,idp1,nonce)) 
    ==> 
    event(beginRegPageReq(admin,idp1,nonce)).
    
query admin:host, idp1:host, nonce:nonce; 
    event(endUserReg(admin,idp1,nonce)) 
        ==> 
    event(beginUserReg(admin,idp1,nonce)).
    
query dapp:host, cc:host, nonce:nonce; 
    event(endSndUsrReg(dapp,cc,nonce)) 
        ==> 
    event(beginSndUsrReg(dapp,cc,nonce)).
\end{lstlisting}

\begin{lstlisting}[caption={Idp resolve and login protocol},captionpos=b, frame=tb]

(* Idp resolve and login protocol *)
query user:host, sp1:host, nonce:nonce;
    event(endLoginReq(user,sp1,nonce)) 
    ==> 
    event(beginLoginReq(user,sp1,nonce)).
query user:host, sp1:host, nonce:nonce;
    event(endServReqVal(user,sp1,nonce)) 
    ==>
    event(beginServReqVal(user,sp1,nonce)).
\end{lstlisting}

These are the queries that we will ask with our ProVerif scripts to validate our protocols for secure communication.

\subsubsection{ProVerif Query Results}
\begin{figure} [hbt]
    \centering
    \includegraphics[width=\linewidth]{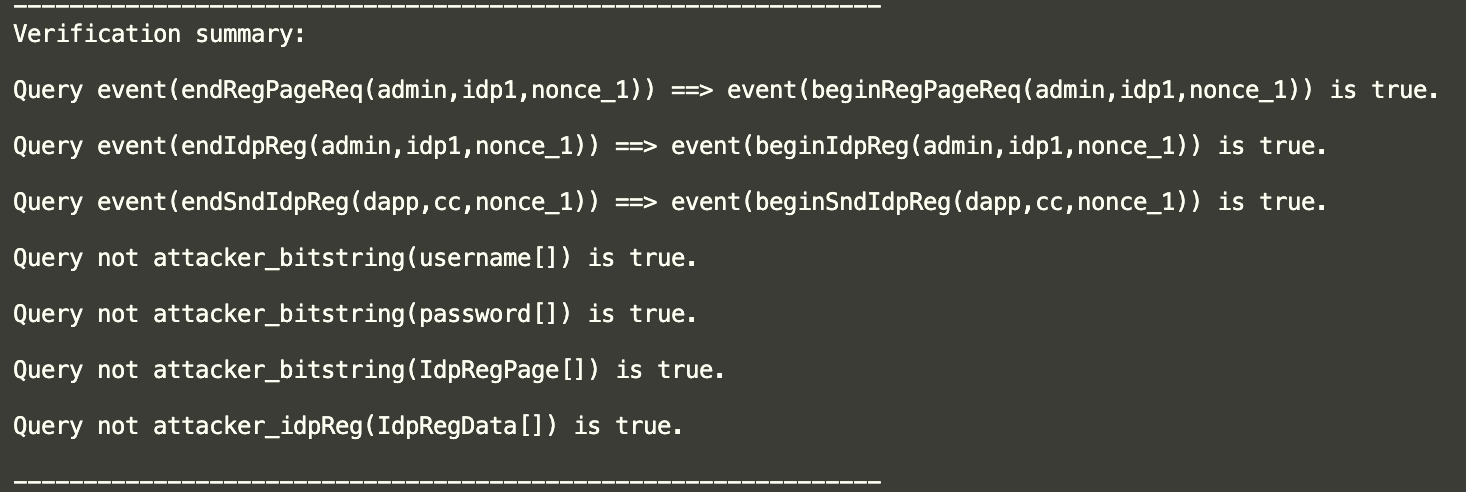}
    \caption{Results of idp registration protocol}
    \label{fig:idpReg}
 \end{figure}

\begin{figure} [hbt]
    \centering
    \includegraphics[width=\linewidth]{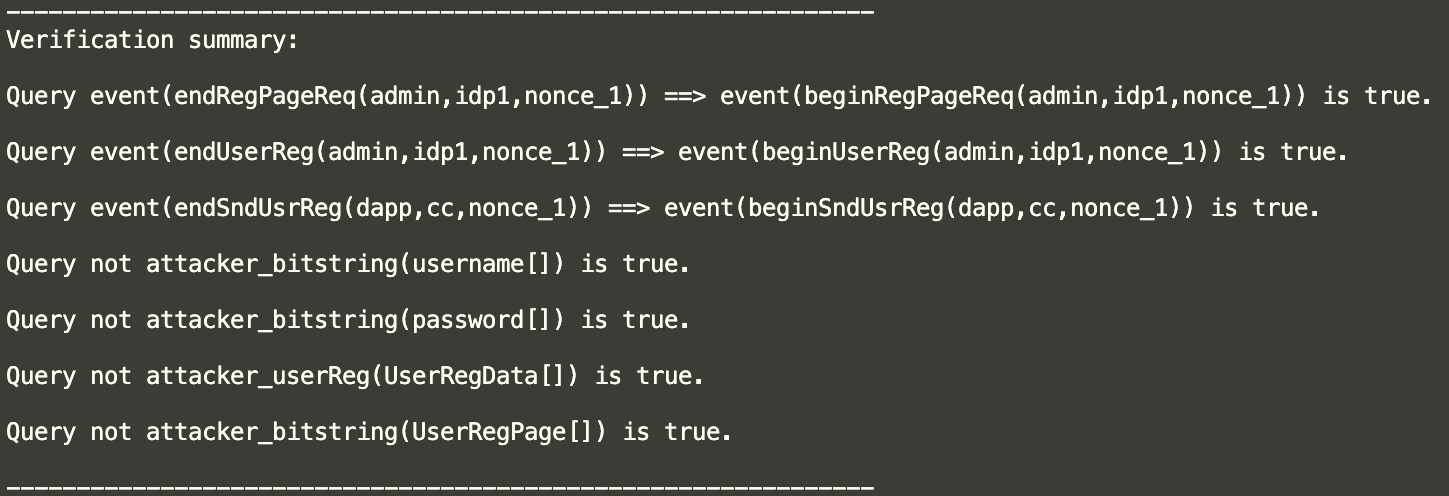}
    \caption{Results of user registration protocol}
    \label{fig:usrReg}
 \end{figure}

\begin{figure} [hbt]
    \centering
    \includegraphics[width=\linewidth]{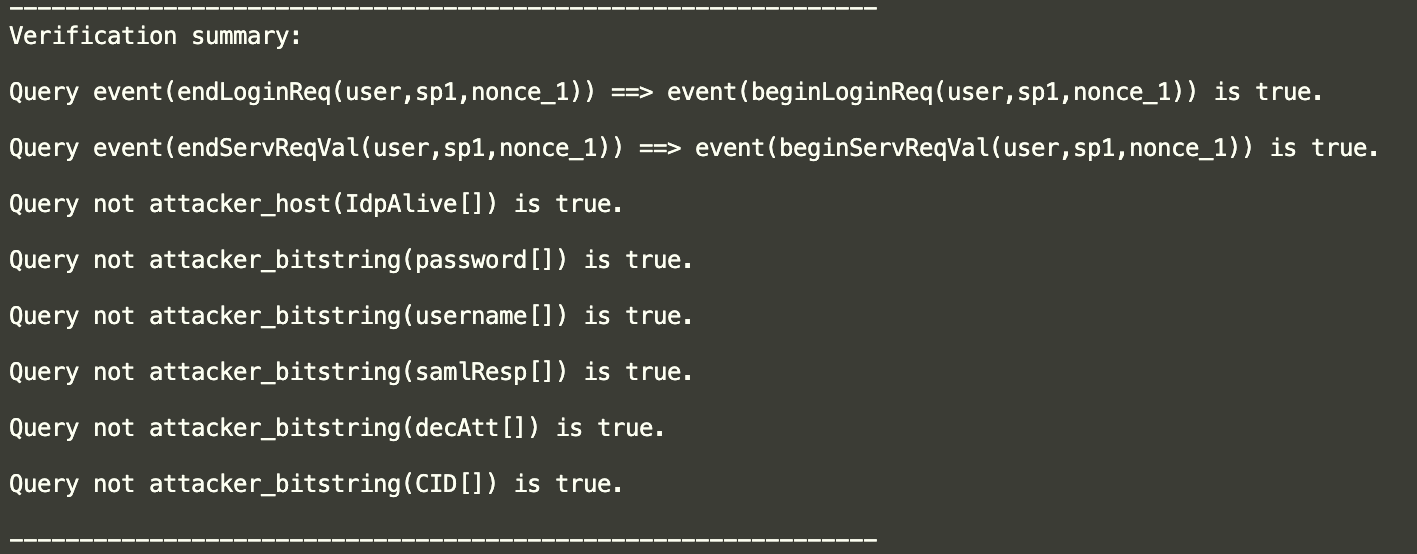}
    \caption{Results of idp resolve and login protocol}
    \label{fig:idpRslv}
 \end{figure}

We will now discuss the findings from the aforementioned queries with the screenshots of the execution of the scripts with the ProVerif executable. These screenshots will confirm that the secrecy and authenticity objectives outlined before have been fulfilled. These screenshots are presented in Figure \ref{fig:idpReg}, Figure \ref{fig:usrReg} and Figure \ref{fig:idpRslv}.



 

Figure \ref{fig:usrReg} simulates the validation result for protocols described in Table \ref{table:userRegProtocol}. We have also demonstrated the protocol verification results for the protocol described in Table \ref{table:regProtocol} in Figure \ref{fig:idpReg} and Protocols described in Tables \ref{table:idpResolvProtocol} and \ref{table:loginProtocol} in Figure \ref{fig:idpRslv}. ProVerif tries to prove the negation of our secrecy reachability goal. According to the results, the secrets are not reachable by an attacker, thereby validating our secreacy goals. ProVerif also corroborates the authenticity of our protocol with injective correspondence, the results are also shown in the corresponding figures.


\subsection{Advantages \& Future Work}
\label{subsec:advantages}
The proposed system offers a number of advantages. Here, we highlight a few of these advantages.
\begin{itemize}
    \item This is the first work to showcase how the blockchain can be leveraged to decentralise an SAML-based identity federation. This ensures that the SPs in this federation will be able to access IdP services even if an IdP ceases to function. 
    \item The system is designed in such a way that the functionalities of SPs within the federation are minimally impacted. For the PoC implementation, we only required to make the majority of changes in the IdP codebase. This is to ensure that the implemented solution can be deployed as easily as possible.
    \item The proposed solution utilises the blockchain platform as an attribute store, thereby providing the immutability of user attributes. In addition, other important information (e.g. CID) is also stored in the blockchain platform. This is an improvement over the traditional SAML implementation where mainly a centralised database is used to store user attributed. 
\end{itemize}

The current version of the implementation only uses smart-contracts (Fabric chaincode) for storing/retrieving information to/from the Fabric blockchain. There are potential advanced use-cases of smart-contracts (e.g. smart-contract based access control within the SAML federations) which have not been explored. In future we would like to explore how such access control mechanism could be integrated within the solution. 

\section{Related Work}
Even though the concept of utilising blockchain within an identity federation is novel, there have been a very few research works in this topic.

ElGayyar et al. presented an idea of a blockchain-based identity federation in \cite{elgayyar2020blockchain}. Their idea is to create a completely new of type of an identity federation marketplace where different entities such as IdPs and SPs could be considered federated, that is within the same circle of trust, if they are part of the same blockchain network. They implemented a prototype based on their proposal leveraging two different blockchain platforms: Ethereum private network and Hyperledger Fabric network. Conceptually the idea presented in this article is similar to the idea presented in \cite{elgayyar2020blockchain}. However, there is one major difference: our concept is based on the SAML protocol where we modified the SAML components to integrate the blockchain network so as to create a decentralised federation. This is to ensure that the existing SAML-based SPs are minimally affected. On the other hand, the work presented in \cite{elgayyar2020blockchain} presents a completely novel identity federation system which does not use SAML. Creating a new protocol and standard would require a significant security analysis based on a rigorous threat model which is completely missing in \cite{elgayyar2020blockchain}.

Another paper related to our concept was presented by Alom et al. in \cite{alom2021dynamic}. In their work, the authors leveraged the Hyperledger Fabric blockchain platform for creating and managing a SAML-based identity federation in a dynamic fashion. The authors did not explore how a SAML federation could be fully decentralised. 

Liu et al. surveyed different blockchain based identity management systems in \cite{liu2020blockchain}. However, they did not look at the possibility of decentralising an identity federation using blockchain and it can be considered as loosely relevant.

\section{Conclusion}
\label{sec:conclusion}
In this article, we have presented a proposal for a novel decentralised identity federation system. The existing SAML-based identity federation systems suffer from a serious centralisation issue because they rely on a single IdP within a federation. This leads to a single point of failure. The proposed blockchain-based decentralised system can effectively tackle this important issue. We have presented its architecture which is based on a rigorous threat model and requirement analysis. We have also designed the protocol flow and developed a Proof-of-Concept (PoC) of the proposed system and showcased several use cases to show the applicability of the developed PoC. We have analysed its performance using several configurations with different user loads. The result of performance testing has yielded Satisfactory results, illustrating the practicality of the system. In addition, we have successfully analysed the security of the underlying protocol of the system using a state-of-the-art protocol verified called ProVerif. 

The concept of decentralisation in identity management is still an active research topic and is likely to be so for years to come. We believe that a decentralised federated identity system could offer many advantages over the traditional federated system. We  hope that our implemented system will be a stepping stone for a future revolutionary change in the Identity Management domain.

\printcredits

\bibliographystyle{cas-model2-names}

\bibliography{main}



\end{document}